\documentclass[11pt,onecolumn,final,oneside]{ar2e}

\usepackage{epsfig}
\usepackage{amssymb}
\usepackage{color}
\usepackage{url}
\usepackage[colorlinks]{hyperref}
\usepackage[italian,american]{babel}
\def\hyref#1{\hyperlink{#1}{#1}}
\begin{document}



\def\ifm#1{\relax\ifmmode#1\else$#1$\fi}  
\def\DAF{DA\char8NE}
\def\x{\ifm{\times}}  
\def\pt#1,#2,{\ifm{#1\x10^{#2}}}
\def\up#1{\ifm{^{#1}}}  
\def\dn#1{\ifm{_{#1}}}
\def\ab{\ifm{\sim}}  
\def\deg{\ifm{^\circ}}  
\def\gam{\ifm{\gamma}} 
\def\to{\ifm{\rightarrow}} 
\def\kl{\ifm{K_L}}   
\def\ks{\ifm{K_S}}
\def\kb{\ifm{\rlap{\kern.3em\raise1.9ex\hbox to.6em{\hrulefill}} K}}
\def\bN{\ifm{\rlap{\kern.3em\raise1.9ex\hbox to.6em{\hrulefill}} N}}
\def\ko{\ifm{K^0}}
\def\kob{\ifm{\kb\vphantom{K}^0}}
\def\kp{\ifm{K^+}}
\def\km{\ifm{K^-}}
\def\eiii{\ifm{\pi^\pm e^\mp\nu}}  
\def\keiii{\ifm{K_{e3}}}
\def\muiii{\ifm{\pi^\pm \mu^\mp\nu}}   
\def\kmuiii{\ifm{K_{\mu3}}}
\def\pio{\ifm{\pi^0\pi^0}} 
\def\po{\ifm{\pi^0}}
\def\pic{\ifm{\pi^+\pi^-}}  
\def\K{\ifm{K}}
\def\rmk{\rm\kern.5mm }   
\def\f{\ifm{\phi}}  
\def\vst{\vphantom{\vrule height4mm depth0pt}}

\def\figbox#1;#2;{\parbox{#2cm}{\epsfig{file=#1.eps,width=#2cm}}}

\newcommand{\MeVc}{\ensuremath{{\rm MeV}}}
\newcommand{\MeVcc}{\ensuremath{{\rm MeV}}}
\newcommand{\GeVcc}{\ensuremath{{\rm GeV}}}
\newcommand{\Lcms}{\ensuremath{{\rm cm}^{-2}\,{\rm s}^{-1}}}
\newcommand{\Lpb}{\ensuremath{\rm pb^{-1}}}

\newcommand{\bra}[1]{\ensuremath{\langle\,#1\,|}}
\newcommand{\ket}[1]{\ensuremath{|\,#1\,\rangle}}
\newcommand{\braket}[2]{\ensuremath{\langle\,#1\,|\,#2\,\rangle}}
\newcommand{\SN}[2]{\ensuremath{#1\times10^{#2}}}

\newcommand{\Eq}[1]{Equation~\ref{#1}}
\newcommand{\Eqs}[1]{Equations~\ref{#1}}
\newcommand{\Fig}[1]{Figure~\ref{#1}}
\newcommand{\Ref}[1]{Reference~\citen{#1}}
\newcommand{\Refs}[1]{References~\citen{#1}}
\newcommand{\Sec}[1]{Section~\ref{#1}}
\newcommand{\Secs}[1]{Sections~\ref{#1}}
\newcommand{\Tab}[1]{Table~\ref{#1}}

\newcommand{\red}{\color{red}}
\newcommand{\blue}{\color{blue}}

\newcommand{\C}{\ensuremath{C}}
\renewcommand{\P}{\ensuremath{P}}
\newcommand{\T}{\ensuremath{T}}
\newcommand{\CP}{\ensuremath{CP}}
\newcommand{\CPT}{\ensuremath{CPT}}

\renewcommand{\Re}[1]{\ensuremath{{\rm Re}\:#1}}
\renewcommand{\Im}[1]{\ensuremath{{\rm Im}\:#1}}
\newcommand{\Reps}{\ensuremath{{\rm Re}\,\epsilon'/\epsilon}}

\newcommand{\Rp}{\ensuremath{R_{\pi}}}
\newcommand{\Rh}{\ensuremath{R_{\eta'}}}

\newcommand{\ita}{\selectlanguage{italian}}


\catcode`@=11 

\newdimen\z@ \z@=0pt 
\newskip\z@skip \z@skip=0pt plus0pt minus0pt
\def\m@th{\mathsurround=\z@}
\def\ialign{\everycr{}\tabskip\z@skip\halign} 
\def\eqalign#1{\null\,\vcenter{\openup\jot\m@th
  \ialign{\strut\hfil$\displaystyle{##}$&$\displaystyle{{}##}$\hfil
      \crcr#1\crcr}}\,}
\catcode`@=12 
\catcode`@=11 
\newcount\equanumber         \equanumber=0
\newdimen\referenceminspace  \referenceminspace=5pc
\let\cl=\centerline

\newcount\figurecount     \figurecount=0
\def\FIG#1{ \global\advance\figurecount by 1 \xdef#1{\the\figurecount}}
\newcounter{tablecntr}
\def\TAB#1{\stepcounter{tablecntr} \xdef#1{\kern.1ex\thetablecntr}}
\def\tblno#1{\hypertarget{#1}{\vglue1pt\cl{\captionfont\bf Table #1}\vglue1pt}}

\def\figboxc#1;#2;{\vglue2mm\cl{\figbox#1;#2;}\vglue3mm}
\def\captionfont{\small}
\let\cl=\centerline
\def\allcap#1;#2;{{\renewcommand{\baselinestretch}{.9}\captionfont
\newdimen\fcwidth \fcwidth=\textwidth \advance\fcwidth by -2cm
\setbox0=\hbox{{\bf Fig. #1.} #2}  
  \ifdim \wd0>\fcwidth  
       \vbox{\noindent
          \parshape=1 1truecm \fcwidth {\bf Fig. #1.} #2}
    \else
       \cl{{\bf Fig. #1.} #2}
    \fi}\renewcommand{\baselinestretch}{1.}\normalsize\hypertarget{#1}\vglue2mm }

\def\to{\ifm{\rightarrow}} \def\sig{\ifm{\sigma}}   \def\plm{\ifm{\pm}}
\def\K{\ifm{K}} \def\LK{\ifm{L_K}} 
\def\ff{$\phi$--factory}
\def\pb{{\bf p}}
\def\eps{\ifm{\epsilon}} \def\epm{\ifm{e^+e^-}}
\def\kpm{\ifm{K^\pm}}  
\def\gam{\ifm{\gamma}}
\def\dt{ \ifm{{\rm d}t} } \def\ab{\ifm{\sim}}  \def\x{\ifm{\times}}
\def\L{\ifm{{\cal L}}}  \def\R{\ifm{{\cal R}}}
\def\pt#1,#2,{\ifm{#1\x10^{#2}}}
\def\vst{\vphantom{\vrule height4mm depth0pt}}
\def\vsta{\vphantom{\vrule height5mm depth0pt}}
\font\euler=eufm10 at 12pt
\def\Ma{\hbox{\euler M}}
\def\ord#1;{\ifm{{\mathcal O}(#1)}}


\renewcommand\section{\@startsection{section}{1}{\z@}%
                                    {-3.5ex \@plus -1ex \@minus -.2ex}%
                                    {2.3ex \@plus.2ex}%
                                    {\reset@font\large\bfseries\mathversion{bold}}}
\renewcommand\subsection{\@startsection{subsection}{2}{\z@}%
                                       {-3.25ex\@plus -1ex \@minus -.2ex}%
                                       {1.5ex \@plus .2ex}%
                                       {\reset@font\large\bfseries\mathversion{bold}}}
\renewcommand\subsubsection{\@startsection{subsubsection}{3}{\z@}%
                                          {.5ex \@plus .2ex}
                                          {-1.5em}
                                          {\reset@font\normalsize\sc}}

\renewcommand\paragraph{\@startsection{paragraph}{4}{\z@}%
                                          {.5ex \@plus .2ex}
                                    {-1em}%
                                    {\reset@font\normalsize\sc}}


\title{The Physics of \DAF\ and KLOE}
\thispagestyle{empty}
\markboth{}{{\rm Franzini \& Moulson} \ The Physics of \DAF\ and KLOE}
\author{
Paolo Franzini
\affiliation{Dipartimento di Fisica dell'Universit\`a 
di Roma, La Sapienza e Sezione dell'Istituto Nazionale di 
Fisica Nucleare, Roma, Italy; email: Paolo.Franzini@lnf.infn.it}
Matthew Moulson
\affiliation{Laboratori Nazionali di Frascati dell'Istituto Nazionale
di Fisica Nucleare, Frascati, Italy; email: Matthew.Moulson@lnf.infn.it}}


\begin{abstract}
\DAF, the Frascati \f\ factory, has been in operation since 1999.
At the center of the physics program is the KLOE experiment, a
multipurpose detector with optimizations for tagged and 
interferometry-based measurements of the neutral kaon system.
KLOE has been taking data since 2000 and has helped to explore
a wide array of topics in kaon and hadronic physics, including
a comprehensive set of measurements to determine the
CKM matrix element $|V_{us}|$, and a measurement of the 
$\epm\to\pic$ cross section for the determination of the
hadronic contribution to the muon anomaly. 
In addition, the DEAR experiment has measured the X-ray spectrum of
kaonic hydrogen, and the FINUDA experiment has conducted its first
studies in hypernuclear spectroscopy and a search for $\kb$-nuclear
bound states. 
We review the design, construction, and
operation of the \DAF\ facility, with an emphasis on the 
physics program of the KLOE experiment.
\end{abstract}

\maketitle

\section{INTRODUCTION}
\label{sec:intro}

The experimental study of kaon decays has played a singular role in 
propelling the development of particle physics for nearly 60 years.
Of particular importance, the 1963 observation at Brookhaven 
National Laboratory of the
decay $\kl\to\pic$ \cite{C+64} was the first experimental evidence of
\CP\ violation in any physical system.
For more than 50 years, the question remained of whether \CP\
violation was confined to the $|\Delta S|=2$ 
$K^0\rightleftarrows\overline K^0$ transition, resulting in a
small \CP\ impurity in the neutral kaon mass
eigenstates, or whether
\CP\ is also violated directly, i.e., in the 
$|\Delta S| = 1$ kaon decay amplitudes.
The former possibility, often referred to as the 
superweak hypothesis \cite{Wol64}, remained viable until quite recently. 
From 1988--1992, the first hints of direct \CP\ violation were
observed \cite{earlyCP}. It soon became clear, however, that
to definitively establish its existence would require increased 
sensitivity.

The search for direct \CP\ violation was the major impetus for the
construction in Frascati of an \epm\ collider to operate at a center-of-mass 
energy equal to the mass of the \f\ meson.
In 1990 the Istituto Nazionale di Fisica Nucleare approved the 
proposal \cite{DAFNE+90:prop} for the collider, named \DAF\
(Double Annular \char8\ factory for Nice Experiments).
The \DAF\ luminosity goal was \SN{5}{32}~\Lcms; that is,
\DAF\ was to produce \ab1300 kaon pairs per
second, or \ab\SN{4}{10} pairs per year at 100\% efficiency.

A first-round search for direct \CP\ violation at \DAF\ would require
an integrated luminosity of \ab10 fb\up{-1}.
The physics program was to be complemented by other topics.
The KLOE experiment \cite{KLOE+92:prop}, named after Chlo\"e, the 
literary counterpart to Daphnis, was proposed in 1992 to 
carry out an ambitious and wide-ranging program
in neutral kaon physics (using especially \ks\kl\ interferometry),
charged kaon physics, and measurements of the properties of scalar 
and pseudoscalar mesons.
The precise determinations of \f\ and $K$ meson properties---masses,
leptonic widths, branching ratios (BRs), and lifetimes---were central to
this program.
In particular, the partial widths for leptonic and semileptonic 
kaon decays are fundamental parameters for comparison with Standard Model
predictions.
Another important element of the program was the measurement of the
hadronic cross section, especially for $\epm\to\pic$, which is
needed to calculate the hadronic contribution to the photon
spectral function in computing the muon anomaly, $(g_\mu-2)/2$.

Two additional experiments were subsequently proposed
to capitalize on the availability of low-energy charged kaons
at the second \DAF\ interaction point: FINUDA 
(FIsica NUcleare a \DAF) \cite{FINUDA+93:prop}
in 1993 for the study of hypernuclear spectroscopy, and 
DEAR (\DAF\ Exotic Atom Research) \cite{DEAR+95:prop} in 1995 
for the spectroscopy of $K$-mesic hydrogen atoms.

In 2005, the \DAF\ luminosity reached \SN{1.3}{32}~\Lcms.
An integrated luminosity of \ab2.5 fb\up{-1} was delivered
to KLOE between 1999 and 2005.
Although far smaller than necessary for a
competitive search for direct \CP\ violation, this data set has permitted
KLOE to explore the breadth of its original physics agenda.
KLOE has performed measurements of the partial decay rates for kaons 
fundamental to the study of quark mixing, has vastly improved 
experimental knowledge of scalar and pseudoscalar meson states,
and has completed a first round of measurements of 
$\sig(\epm\to {\rm hadrons})$.
In addition, DEAR has measured the level shifts and widths
of kaonic hydrogen, and FINUDA has obtained first results on hypernuclear
spectroscopy and conducted a search for new states from nuclear \km\
absorption.

\FIG\dafnelay   \FIG\intLum
\section{\DAF: THE FRASCATI \f\ FACTORY}

\DAF, like PEP-II at the Stanford Linear Accelerator Center and KEKB
at the KEK laboratory in Japan, uses two separate rings to store
large numbers of electron and positron bunches, thus avoiding beam-beam 
interaction effects.
The \DAF\ design parameters---a luminosity goal of \SN{5}{32}~\Lcms\
in 120-bunch operation in a machine with rings 98~m 
in circumference---imply a stored current of approximately 5~A in each beam.
The beams intersect at an angle of $(\pi - 0.025)$~rad in two interaction 
regions (IRs) on opposite sides of the rings, where low-$\beta$ 
insertions provide focusing to obtain high luminosity \cite{B+91}.
At the IRs, the bunches are flat in cross section, with 
$\sigma_x \approx 2$~mm and 
$\sigma_y \approx 10~\mu$m.
Figure \hyref{\dafnelay} shows a plan view of the \DAF\ complex.
\\\figboxc NS56pre_fig01;14;
\allcap\dafnelay;Schematic diagram of the \DAF\ facility.;\noindent
A two-stage linac produces electrons and positrons of 510~MeV, with a
repetition rate of 50~Hz. These are stacked and cooled 
in the accumulator ring.
Bunches are transferred from the accumulator to the \DAF\ rings
on-energy at a rate of 1~Hz.
Each \DAF\ ring consists of four bending sections, each with a wiggler 
magnet, and a large number of quadrupoles to provide the basic 
alternating-gradient focusing structure.
The four wigglers in each ring help to reduce the 
damping times for synchrotron and betatron oscillations, which still
remain \ab50,000 times larger than the cyclotron period.

\DAF\ operations got off to a rough start.
The highest luminosity sustained during the year following KLOE roll-in
in April 1999 was \ab\SN{2}{30}~\Lcms\ with 40 bunches, or \ab2\% of the 
day-one target value. 
Since then, the luminosity has been raised slowly by 
abandoning attempts to provide simultaneous collisions at both IRs,
adding octupole magnets, correcting the poor wiggler field shape,
improving injection efficiency,
and installing a redesigned low-$\beta$ insertion at the KLOE IR.
Even the KLOE field setting has been adjusted to optimize the 
luminosity \cite{Zob00}. 
The integrated luminosity is maximized by frequently topping up
the stored beams, without interrupting KLOE data taking.
In 2000, an integrated luminosity of 24~pb$^{-1}$ was delivered.
This data set was used
to obtain the first generation of KLOE results.

\DAF\ performance improved considerably year by year from 2001 to 2005.
Figure \hyref{\intLum} illustrates the growth of the KLOE data set.
\\\figboxc NS56pre_fig02;14;
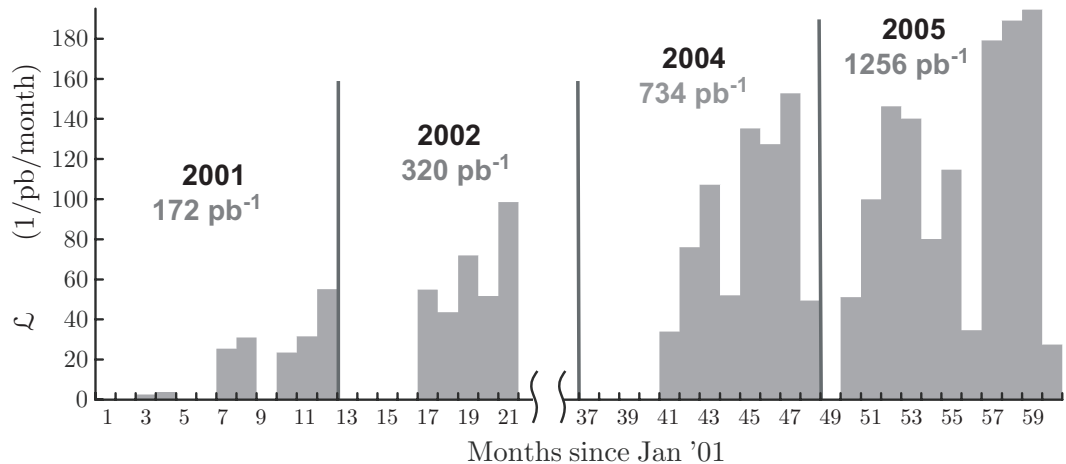
\allcap\intLum;Luminosity integrated by KLOE from 2001 to 2005.;\noindent
In 2002, 107~pb$^{-1}$ were delivered to DEAR.
The first physics run with FINUDA began in November 2003 and ended in
March 2004; a total of 250~pb$^{-1}$ were delivered.   
From May 2004 to December 2005, \DAF\ operations 
were dedicated to KLOE data taking.
In a single month near the end of this period (November 2005), 
KLOE collected 190~pb$^{-1}$, with
sustained luminosities regularly in excess of $10^{32}$~\Lcms\
and a one-day integrated-luminosity record of 8.8~\Lpb.
During the 2004--2005 run, KLOE collected 1.99~fb$^{-1}$.

\FIG\kloesec
\section{THE KLOE EXPERIMENT}
\label{sec:kloe}

At \DAF, \f\ mesons decay nearly at rest.\footnote
{Because of the beam-crossing angle, \f s have a net momentum of 
12--16~\MeVc\ in the horizontal plane and directed towards the center
of the rings. This momentum is determined run by run from Bhabha scattering
events with a precision of 0.01~\MeVc.} 
Neutral and charged kaons from \f\ decays have momenta of 110 and 127~\MeVc,
respectively. 
As a result, the mean \kl, \ks, and \kpm\ decay path lengths
are $\lambda_L = 3.4$~m, $\lambda_S = 0.59$~cm, and $\lambda_\pm = 95$~cm.
A detector with a radius of \ab2~m is required to 
define a fiducial volume for the detection of \kl\ decays with a 
geometrical efficiency of \ab30\%.
Because the radial distribution of the \kl\ decay points is
essentially uniform within this volume, tracks must be well reconstructed
independently of their angles of emission, and photon vertices
must be localized.
To observe $K_S$ decays and $K_SK_L$ interference patterns
with minimal complications from $K_L\to K_S$ regeneration, a decay 
volume about the interaction point with $r \gtrsim 10\lambda_S$ must 
remain in vacuum.
Material within the sensitive volume must be kept
to a minimum to reduce the effects of regeneration, photon conversion,
and multiple scattering and energy loss for low-momentum charged particles.

Figure \hyref{\kloesec} shows a diagram of the cross section of the KLOE detector.
The detector consists principally of a large drift chamber (DC) surrounded 
by a hermetic electromagnetic calorimeter (EMC). A superconducting coil
surrounding the calorimeter provides an axial magnetic field of 0.52~T.
\\\figboxc NS56pre_fig03;12;
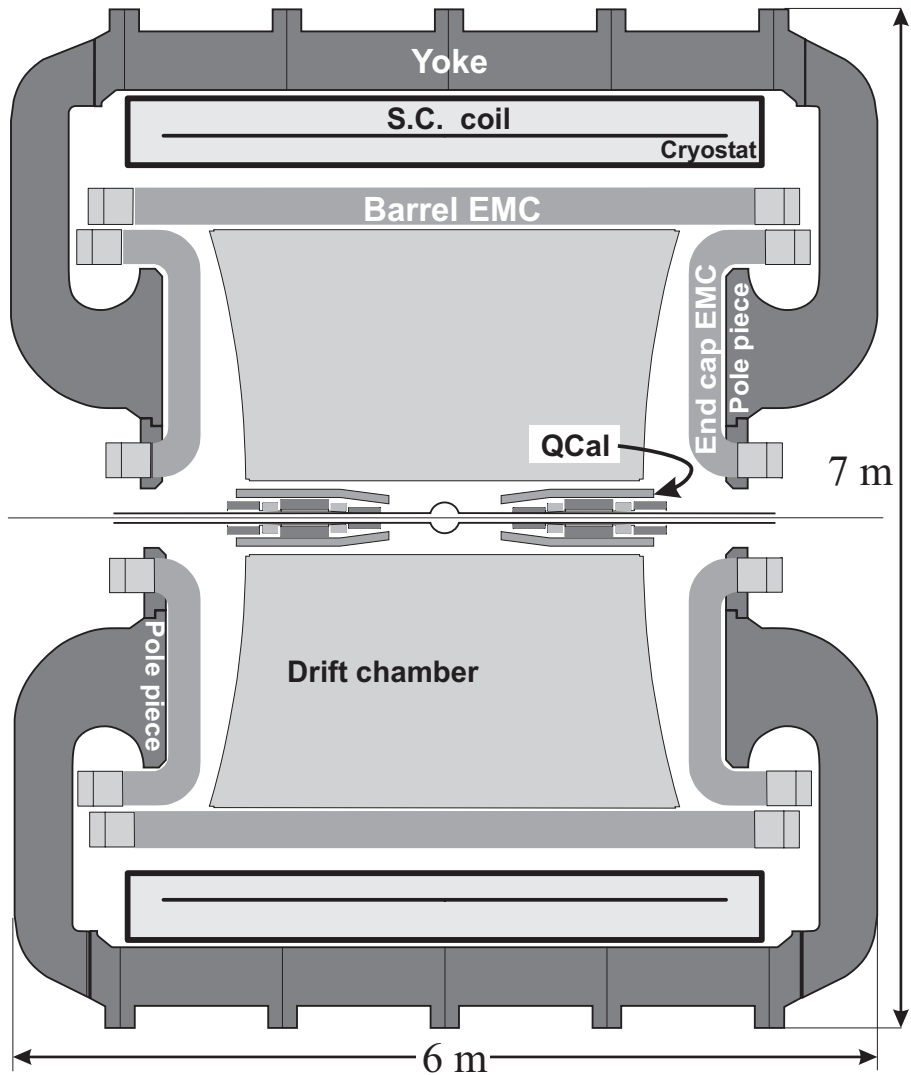
\allcap\kloesec;Cross-sectional view of the KLOE experiment, showing the interaction region, the drift chamber (DC), the electromagnetic calorimeter (EMC), the superconducting coil, and the return yoke of the magnet.;

The DC is 3.3~m long, with inner and outer radii of~25 
and 200~cm, respectively. It contains 12,582 drift cells arranged in
58 stereo layers, for a total of 52,140 wires.
The absolute value of the stereo angle increases from 60 to 150~mrad 
with the layer radius, and the sign of the stereo angle alternates
from layer to layer. 
The difference between the radial distance from the chamber axis to the 
wires on a given layer as measured at the endplates and at the 
midpoint of the chamber is the same for all layers (1.5~cm).
This design results in uniform filling of the sensitive
volume, which increases the isotropy of the tracking efficiency.
The cells are approximately square in shape; those in the
first 12 layers measure $2\times2~{\rm cm}^{2}$ in cross section, 
whereas those in the remaining 46 layers measure $3\times3~{\rm cm}^2$.
The chamber uses a gas mixture of 90\% helium and 10\% isobutane.
This reduces regeneration and multiple scattering within the 
chamber, while providing good spatial resolution (150~$\mu$m).
Large-angle tracks from the origin ($\theta > 45\deg$) are 
reconstructed with $\sigma_p/p < 0.4\%$, and vertices within 
the sensitive volume are reconstructed with a 
position resolution of \ab3~mm.
Groups of 12 consecutive wires on each layer are read out by
analog-to-digital-converter electronics;
the measurement of the specific ionization allows identification of
\kpm\ tracks by $dE/dx$ alone. 
A full description of the design and operation of the chamber can
be found in \Ref{DCNIM}.

The calorimeter is made of cladded, 1-mm scintillating 
fibers sandwiched between 0.5-mm-thick lead foils.
The foils are imprinted with grooves wide enough to accommodate 
the fibers and some epoxy, without compressing the fibers.
This precaution prevents damage to the fiber-cladding interface.
The epoxy around the fibers also provides structural strength
and removes light traveling in the cladding. 
Many such layers are stacked, glued, and pressed, resulting
in a bulk material with a radiation length $X_0$ of 1.5~cm
and an electromagnetic sampling fraction of \ab13\%.
This material is fashioned into modules 23~cm thick (\ab$15X_0$),
24 of which are arranged in azimuth to form the calorimeter barrel,
and an additional 32 of which are wrapped around each of the
pole pieces of the magnet yoke to form the endcaps.
The unobstructed solid-angle coverage of the calorimeter
as viewed from the origin is \ab94\%. 
The fibers run parallel to the axis of the detector in the barrel,
run vertical in the endcaps, and are read out at both ends 
with a granularity of $4.4\times4.4~{\rm cm}^2$ by a total of 4880 
photomultiplier tubes. 
Cluster energies are measured with a resolution of
$\sigma_E/E = 5.7\%/\sqrt{E\ {\rm(GeV)}}$, as determined with the help of
the DC using radiative Bhabha events.
The absolute time resolution is $\sigma_t = 54~{\rm ps}/\sqrt{E\ {\rm (GeV)}}
\oplus 140~{\rm ps}$, as determined from radiative \f\ decays. The constant
term results largely from the uncertainty of the event $t_0$ arising from
the length of the \DAF\ bunches.
The constant contribution to the relative time resolution as 
determined using $2\gamma$ events is \ab50~ps.  
Cluster positions are measured with resolutions of 1.3~cm in the 
coordinate transverse to the fibers, and,
by timing, of
$1.2~{\rm cm}/\sqrt{E\ {\rm(GeV)}}$ in the longitudinal coordinate.
These characteristics enable the $2\gamma$ vertex in 
$\kl\to\pic\po$ decays to be localized with $\sigma \approx 2$~cm 
along the \kl\ line of flight, as reconstructed from the tagging \ks\
decay. The calorimeter is more fully described in \Ref{EmCNIM}.

Around the interaction point, the beam pipe is spherical in shape,
with a radius of 10~cm, so that all $K_S$ mesons decay in vacuum.
The beam-pipe walls are made of a 60\%-beryllium/40\%-aluminum
alloy 0.5~mm thick.
The quadrupoles of the low-$\beta$ insertion are covered with a
lead/scintillating-tile calorimeter \cite{QNIM} intended to
detect photons that are otherwise absorbed on the quadrupoles.

The two-level KLOE trigger \cite{trigNIM} uses information from both 
the calorimeter and the DC. The level-1 trigger provides
a fast response to initiate conversion in the front-end electronics
modules. It is satisfied by 
two energy deposits above a threshold of 50~MeV on the EMC
barrel and above 150~MeV on the endcaps, or by \ab15 hits in the DC 
arriving within a time window of 250~ns.
All detector signals are digitized by KLOE-designed modules
that can operate on signals arriving before the level-1 trigger.
The level-2 trigger 
initiates event read out. In a typical configuration, the level-1 
EMC signal satisfies level 2, whereas for the DC, \ab120 hits must 
arrive within a 1.2-$\mu$s window.
The trigger also implements logic to identify cosmic-ray events,
which are recognized by the presence of two energy deposits
above 30~MeV in the outermost calorimeter plane. 

At a luminosity of $10^{32}$~\Lcms, events are recorded at \ab2200~Hz.
Of this rate, \ab300~Hz are from \f\ decays.
Raw data, reconstructed data, and Monte Carlo (MC) events occupy 
\ab800~TB and are stored in a tape library. Data summary tapes for
both data and MC occupy \ab80~TB and are cached on disk for analysis.
For a detailed description of the data acquisition, online, and offline
systems, see \Refs{DAQNIM} and \citen{offlineNIM}.

\section{KAON PHYSICS}

\subsection{Why a \f\ Factory?}
\label{sec:ff}
Interest in a \f\ factory \cite{DHR87,DAFNE91,ref:paolo}
is due to the fact that \f\ mesons are produced abundantly
in \epm collisions. The visible cross section for $\epm\to\f$ peaks at
\ab3~$\mu$b at $\sqrt s \approx 1019.4$~MeV\@.
For comparison, $\sigma(\epm\to{\rm hadrons}) \approx 0.1$~$\mu$b,
and $\sigma(\epm\to\epm,\ \theta>20\deg) \approx 7$~$\mu$b.
The \f\ meson decays dominantly to charged kaon pairs (49\%), neutral kaon
pairs (34\%), $\rho\pi$ (15\%), and $\eta\gam$ (1.3\%).
A \f\ factory is thus a copious source of tagged and monochromatic
kaons, both neutral and charged, and even a source of $\eta$ mesons.

\subsubsection{Tagged Kaon Beams}
Because the neutral kaon pair from $\f\to\ko\kob$
is purely $J^{PC}$=1\up{--}, the initial two-kaon state can be written
as
\begin{equation}
\ket{K\kb,\ t=0} = (\ket{\ko\kob}-\ket{\kob\ko})/\sqrt{2}
\equiv(\ket{\ks\kl} - \ket{\kl\ks})/\sqrt{2},
\label{eq:kkp}
\end{equation}
where the identity holds even without assuming \CPT\ invariance. Detection
of a \ks\ (\kl) thus signals the presence of a \kl\ (\ks).
This in effect creates pure \ks\ and \kl\ beams of precisely known momenta 
(event by event, from kinematic closure) and flux, which can be used to 
measure absolute $K_S$ and $K_L$ BRs, or, with even greater
precision, ratios of such BRs. Similar arguments hold for 
$K^+$ and $K^-$ as well.

\FIG\interf
\subsubsection{\ks\kl\ Interferometry}
\label{sec:inter}
Assuming \CPT\ invariance, to lowest order in \eps,
the states \ket{\ks} and \ket{\kl} are given by \cite{DHR87,ref:paolo} 
\begin{equation}
\ket{\ks}={(1+\eps)\ket{\ko}+(1-\eps)\ket{\kob}\over\sqrt N},\  
\ket{\kl}={(1+\eps)\ket{\ko}-(1-\eps)\ket{\kob}\over\sqrt N},
\label{eq:kslcpt}
\end{equation}
where $N=2(1+|\eps|^2)$. If \CPT\ invariance is not assumed, then ($N$
changes):
\begin{equation}
\ket{\ks}={(1+\eps_S)\ket{\ko}+(1-\eps_S)\ket{\kob}\over\sqrt N},\ 
\ket{\kl}={(1+\eps_L)\ket{\ko}-(1-\eps_L)\ket{\kob}\over\sqrt N}.
\label{eq:kslncpt}
\end{equation}
It is normal to define the parameters $\tilde\eps$ and $\delta$ through
the identities
\begin{equation}
\eps_S\equiv\tilde\eps+\delta;\qquad\eps_L\equiv\tilde\eps-\delta.
\label{eq:epsdel}
\end{equation}
We evolve the initial state in \Eq{eq:kkp} in time, project to any two 
possible final states $f_1$ and $f_2$, take the modulus
squared, and integrate over all $t_1$ and $t_2$ for fixed $\Delta t=t_1-t_2$
to obtain (for $\Delta t > 0$, with $\Gamma\equiv\Gamma_L+\Gamma_S$)
\begin{equation}
\eqalign{
I_{f_1,\,f_2}(\Delta t)&=
{1\over2\Gamma}|\braket{f_1}{\ks}\braket{f_2}{\ks}|^2 \x
\Big[|\eta_1|^2e^{-\Gamma_L\Delta t}+|\eta_2|^2e^{-\Gamma_S\Delta t}\cr
& -2|\eta_1||\eta_2|
e^{-(\Gamma_L+\Gamma_S)\Delta t/2}\cos(\Delta m\Delta t+\f_2-\f_1)\Big].\cr}
\label{eq:eqtwo}
\end{equation}
The last term is due to interference between the decays to states 
$f_1$ and $f_2$.
Fits to the $\Delta t$ distribution provide measurements of the
magnitudes and phases of the parameters 
$\eta_i = \braket{f_i}{\kl}/\braket{f_i}{\ks}$, as well as of the 
\kl-\ks\ mass difference $\Delta m$ and the decay rates 
$\Gamma_L$ and $\Gamma_S$.
Many examples are discussed in \Refs{DHR87}, \citen{B+92:inter},
and \citen{DAIP95:inter}.

Such fits also allow tests of fundamental properties of quantum mechanics.
For example, the persistence of 
quantum-mechanical coherence can be tested by choosing $f_1 = f_2$. 
In this case, because of the antisymmetry 
of the initial state and the symmetry of the final state, there 
should be no events with $\Delta t = 0$. 
Using the 2001--2002 data, KLOE has conducted a preliminary analysis of 
the $\Delta t$ distribution for $\ks\kl\to\pic\pic$ events
that establishes the feasibility of such tests \cite{KLOE+05:inter}.
The $\Delta t$ distribution is fit with a function of the form of 
\Eq{eq:eqtwo}, including the experimental resolution and a peak from
$\kl\to\ks$ regeneration in the beam pipe. The results are shown in
Figure \hyref{\interf}.
\figboxc NS56pre_fig04;8;
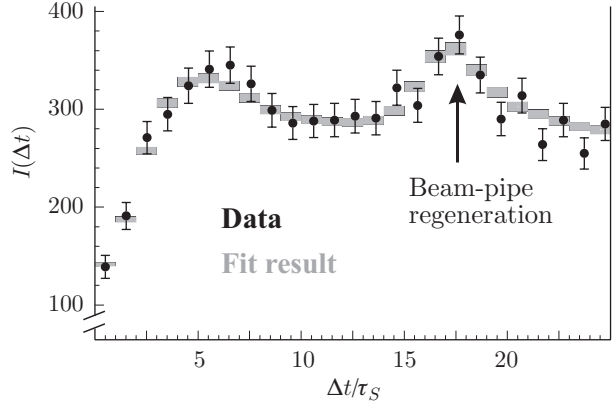
\allcap\interf;$\Delta t$ distribution for $\f\to\ks\kl\to\pic\pic$
events in 380~\Lpb\ of KLOE data.; 

\subsubsection{Direct $CP$ Violation in Neutral Kaon Decays}
\Eq{eq:kslcpt} gives the expressions for the mass eigenstates 
\ket{\ks} and \ket{\kl} assuming \CPT\ invariance.
If \CP\ violation in the \ks\kl\ system results exclusively from mixing,
$\eta_{\pic}\equiv\eta_{+-}=\eta_{\pio}\equiv\eta_{00}=\eps$.
If \CP\ is violated in the $|\Delta S| = 1$ decay amplitudes,
the amplitude ratios $\eta$ are no longer
equal and the magnitude of direct \CP\ violation can be parameterized
by $\eps'$:
\begin{displaymath}
\eta_{+-}=\eps+\eps';\kern1cm\eta_{00}=\eps-2\eps'.
\end{displaymath}
In the Standard Model, both \eps\ and $\eps'$ are calculable in terms of
the Cabibbo-Kobayashi-Maskawa (CKM) matrix elements. 
In general, if \eps\ is due to
\CP\ violation in the CKM matrix, $\eps'\ne0$, except for fortuitous 
cancellations.
There are no reliable calculations of $\eps'$. Experimentally, 
\Reps\ is obtained from the double ratio:
\begin{equation}
{\cal R}\equiv{\Gamma(\kl\to\pio)/\Gamma(\ks\to\pio)\over
\Gamma(\kl\to\pic)/\Gamma(\ks\to\pic)}\equiv
\Big|{\eta_{00}\over\eta_{+-}}\Big|^2=1-6\,\Reps.
\label{eq:rr}
\end{equation}
Recent results for \Reps\ from NA48 at CERN \cite{NA48+02:final} 
and KTeV at Fermilab \cite{KTeV+03:final} are
\SN{(14.7\pm2.2)}{-4} and
\SN{(20.7\pm2.8)}{-4}, respectively.
The two results are not in perfect agreement, but prove
beyond any doubt the existence of 
direct \CP\ violation; the observed value of \Reps\ 
is compatible with general expectations.
The existence of direct \CP\ violation cannot be confirmed at \DAF\
because of lack of luminosity.
With sufficient data, \Reps\ would be measurable by 
\Eq{eq:rr}, and both \Reps\ and \Im{\eps'/\eps}
would be measurable by interferometry.

\TAB\vusw
\subsubsection{Kaon Decays and Determination of $|V_{us}|$}
\label{sec:vus} 
In the Standard Model, the quark weak charged current is
\begin{displaymath}
J_\alpha^+=(\bar u\ \bar c\ \bar t)\gamma_\alpha(1-\gamma_5)\,{\bf V}
\pmatrix{d\cr s\cr b\cr},
\end{displaymath} 
where {\bf V} is a $3\times3$ unitary matrix introduced by
Kobayashi and Maskawa \cite{KM73} in expansion on an original
suggestion by Cabibbo \cite{Cab63}.
The unitarity condition (${\bf V}^\dagger{\bf V}=1$) is required
by the assumption of universality of the weak interactions of leptons
and quarks and the absence of flavor-changing neutral currents.
The realization that a precise test of CKM unitarity can be obtained 
from the first-row constraint
$|V_{ud}|^2 + |V_{us}|^2 + |V_{ub}|^2 = 1$ (with $|V_{ub}|^2$ negligible)
has sparked a new interest in good measurements of quantities related to
$|V_{us}|$.
As we discuss in the following sections, $|V_{us}|$ can be determined 
using semileptonic kaon decays;
the experimental inputs are the BRs, lifetimes, 
and form-factor slopes. Both neutral (\ks\ or \kl) and charged kaons
may be used and provide independent measurements.
Many players have joined the game, as seen from Table \hyref{\vusw}.\vglue2mm
\begin{center}
\cl{Table \vusw. Recent world data on $K_{\ell3}$ decays for calculation of $|V_{us}|$ }\vglue2mm
\begin{tabular}{@{}lll@{}}
\hline\hline
\vst Experiment & Parameters measured & References \\
\hline
\vsta E865 & BR($\kp\to\po_{\rm D}e^+\nu$)/BR($\kp\to\pi^0_{\rm D}X^+$) & \cite{E865+03:Ke3} \\
\vsta KTeV & BR($K_{L\,e3}$), BR($K_{L\,\mu3}$), $\lambda_+(K_{L\,e3})$, $\lambda_{+,0}(K_{L\,\mu3})$ & \cite{KTeV+04:BR,KTeV+04:FF} \\
\vsta ISTRA+ & $\lambda_+(K^-_{e3})$, $\lambda_{+,0}(K^-_{\mu3})$ & \cite{ISTRA+04:Ke3,ISTRA+04:Kmu3} \\
\vsta NA48 & BR($K_{L\,e3}$)/BR(2 tracks), BR($K^\pm_{e3}$)/BR($\pi\po$), $\lambda_+(K_{L\,e3})$ & \cite{NA48+04:Ke3L,NA48+04:ICHEP,NA48+04:Ke3FF} \\
\vsta KLOE & BR($K_{L\,e3}$), BR($K_{L\,\mu3}$), BR($K_{S\,e3}$), BR($K^\pm_{e3}$), BR($K^\pm_{\mu3}$), & See text \\  & $\lambda_+(K_{L\,e3})$, $\tau_L$, $\tau^\pm$ & \\
\hline
\end{tabular}
\end{center}\hypertarget{\vusw}\kern0pt\vglue2mm
\noindent
KLOE is unique in that it is the only experiment that
can by itself measure the complete set of experimental inputs for the
calculation of $|V_{us}|$ using both charged and neutral kaons.
This is because the \f\ factory is uniquely suited for measurements
of the \kl\ and \kpm\ lifetimes.
In addition, KLOE is the only experiment that can measure \ks\
BRs at the sub-percent level.
We illustrate in \Sec{sec:KLOEvus} by calculating $|V_{us}|$ from the 
comprehensive KLOE data set on semileptonic kaon decays.

One problem that consistently plagues the interpretation of older 
BR measurements is lack of clarity in the treatment of
radiative contributions.
All KLOE measurements of kaon decays with charged particles in the 
final state are fully inclusive of radiation. The inclusion of 
radiation is handled as an acceptance correction---the relevant MC
generators incorporate radiation as described in \Ref{Gat05}.

\subsubsection{Semileptonic Kaon Decays}
\label{sec:vusl3}
The semileptonic kaon decay rates
still provide the best means for the measurement of $|V_{us}|$
because only the vector part of the weak current contributes
to the matrix element $\bra{\pi}J_\alpha\ket{K}$. In general,
\begin{displaymath}
\bra{\pi}J_\alpha\ket{K} = f_+(t)(P+p)_\alpha + f_-(t)(P-p)_\alpha,
\end{displaymath}
where $P$ and $p$ are the kaon and pion four-momenta, respectively,
and $t=(P-p)^2$.
The form factors $f_+$ and $f_-$ appear because pions and kaons are 
not point-like particles, and also reflect both $SU(2)$ and $SU(3)$ 
breaking. For vector
transitions, the Ademollo-Gatto theorem \cite{ag} ensures that
$SU(3)$ breaking appears only to second order in $m_s-m_{u,d}$. 
In particular, $f_+(0)$ differs from unity by only 2--4\%. 
When the squared matrix element is evaluated, a factor of $m_\ell^2/m_K^2$
multiplies all terms containing $f_-(t)$. This form factor can be 
neglected for $K_{e3}$ decays. For the description of $K_{\mu3}$ decays,
it is customary to use $f_+(t)$ and 
the scalar form factor $f_0(t) \equiv f_+(t) + [t/(m_K^2-m_\pi^2)]\,f_-(t)$. 

The semileptonic decay rates, fully inclusive of radiation, are given by
\begin{equation}
\Gamma^i(K_{e3,\,\mu3})=|V_{us}|^2\:{C_i^2\:G^2\:M^5\over768\pi^3}\:S_{\rm
EW}\:(1+\delta_{i,\,\rm em}+
\delta_{i,\,SU(2)})\:|f^{\ko}_+(0)|^2\:I_{e3,\,\mu3}.
\label{eq:Gamsl}
\end{equation}
In the above expression, $i$ indexes $\ko\to\pi^\pm$ and $K^\pm\to\po$
transitions, for which $C_i^2 =1$ and 1/2, respectively. $G$ is the Fermi
constant, $M$ is the appropriate kaon mass, and $S_{\rm EW}$ is the
universal short-distance radiative correction factor \cite{as}. 
The $\delta$ terms are the long-distance
radiative corrections, which depend on the meson charges and lepton masses,
and the $SU(2)$-breaking corrections, which depend on the kaon charge 
\cite{aa}. 
The form factors are written as 
$f_{+,\,0}(t)=f_+(0)\tilde f_{+,\,0}(t)$, with $\tilde f_{+,\,0}(0)=1$. 
$f_+(0)$ reflects $SU(2)$- and $SU(3)$-breaking
corrections and is different for $\ko$ and $K^\pm$. $I_{e3,\,\mu3}$
is the integral of the Dalitz-plot density over the physical region
and includes $|\tilde f_{+,\,0}(t)|^2$.  $I_{e3,\,\mu3}$ does not 
account for photon emission; the effects of radiation are included
in the electromagnetic (em) corrections. The numerical factor
in the denominator of \Eq{eq:Gamsl}, $768=3\x2\up8$, is chosen in
such a way that $I=1$ when the masses of all final-state particles vanish.
For $K_{e3}$, $I \approx 0.56$ and for $K_{\mu3}$, $I \approx 0.36$. 
The vector form factor $f_+$ is dominated by the vector $K\pi$ resonances, the
closest being the $K^*(892)$. Note that for $t>0$, $\tilde f_+(t)>1$. The
presence of the form factor increases the value of the phase-space integral
and the decay rate.
The natural form for $\tilde f_+(t)$ is
\begin{equation}
\tilde f_+(t) ={M_V^2\over M_V^2-t}.
\label{eq:pole}
\end{equation}
It is also customary to expand the form factor in powers of $t$ as
\begin{displaymath}
\tilde f_+(t)=1+\lambda'{t\over m^2_{\pi^+}}+
{\lambda''\over2}\left({t\over m^2_{\pi^+}}\right)^2.
\end{displaymath}
To compare the results obtained from each semileptonic decay mode
for both neutral and charged kaons without knowledge of $f_+(0)$,
\Eq{eq:Gamsl} is usually used to compute the quantity $f_+^{K^0}(0)\,|V_{us}|$.
This requires the $SU(2)$ and electromagnetic corrections for all four 
possible cases.

\subsubsection{$K\to\mu\nu$ Decays}
High-precision lattice quantum chromodymanics (QCD) results have 
recently become available and are
rapidly improving \cite{lat}. The availability of precise values for the
pion- and kaon-decay constants $f_\pi$ and $f_K$ allows use of a relation
between $\Gamma(K_{\mu2})/\Gamma(\pi_{\mu2})$ and $|V_{us}|^2/|V_{ud}|^2$,
with the advantage that lattice-scale uncertainties and radiative corrections
largely cancel out in the ratio \cite{ref:marfk}:
\begin{equation}
{\Gamma(K_{\mu2(\gamma)})\over\Gamma(\pi_{\mu2(\gamma)})}=%
{|V_{us}|^2\over|V_{ud}|^2}\;{f_K^2\over f_\pi^2}\;%
{m_K\left(1-m^2_\mu/m^2_K\right)^2\over m_\pi\left(1-m^2_\mu/m^2_\pi\right)^2}\x(0.9930\pm0.0035),
\label{eq:fkfp}
\end{equation}
where the precision of the numerical factor due to structure-dependent 
corrections \cite{ref:fink} can be improved.
Thus, it could very well
be that the abundant decays of pions and kaons to $\mu\nu$ ultimately
give the most accurate determination of the ratio of $|V_{us}|$ to 
$|V_{ud}|$.
This ratio can be combined with direct measurements of $|V_{ud}|$ to obtain
$|V_{us}|$ using unitarity~\cite{ref:marfk,ref:fkfp}.
What is more interesting, however, is to combine all information
from $K_{e2}$, $K_{\mu2}$, $K_{e3}$, $K_{\mu3}$, and superallowed 
$0^+ \to 0^+$ nuclear $\beta$-decays to
experimentally test electron-muon and lepton-quark universality,
in addition to the unitarity of the quark mixing matrix. 

\subsection{$K_L$ Decays}
\label{sec:kldec}

\subsubsection{Measurements of \kl\ properties}
Not all parameters describing the \ks\kl\ system are
measurable from the study of $\f\to\ks\kl$ decays
without assuming either \CPT\ invariance or the $\Delta S=\Delta Q$ rule.
The set can be made complete at \DAF\ by including measurements of the
$\Delta S \neq \Delta Q$ amplitude using strangeness-tagged 
\ko\ and \kob\ mesons
from charge exchange of tagged \kpm\ mesons. We are mostly interested here
in the quantities relevant to the measurement of $|V_{us}|$. 
For \kl\ mesons, the necessary quantities are the neutral kaon mass,
the \kl\ lifetime, the semileptonic BRs, and the slopes  
of the hadronic-current form factors.
KLOE finds $m_{\kl}=497.583\pm0.021$~\MeVcc\ \cite{KLOE+05:llwidth,ref:kmass}.
This value is accurately determined owing to the precise Novosibirsk
measurement of the \f\ mass \cite{CMD2phi} and the smallness of 
$m_\f-2m_{\ko} = 24.317$~\MeVcc. 
Knowledge of the mass is necessary for the 
phase-space integrals, for the BR measurements, for the correct
evaluation of the $M^5$ factor in \Eq{eq:Gamsl}, and for the evaluation of 
the radiative corrections. 

\subsubsection{$K_L$ Lifetime}
\label{sec:KLlife}
The \kl\ lifetime is particularly difficult
to measure because monochromatic neutral kaons are generally not available,
nor can they be stopped. In addition, only a small fraction of the \kl\
lifetime is covered in a typical experiment.
If $N$ events are observed in a time window of $T$ lifetimes (with $T \ll 1$),
and the lifetime is determined from a fit to the proper-time
distribution, the fractional error is $\delta\tau/\tau=2\sqrt3/(T\sqrt N)$. 
A good measurement is possible at \DAF, where a monochromatic beam of 
very slow \kl s is available. KLOE can cover \ab37\% of the \kl\ lifetime, 
compared with a few percent in KTeV or NA48.
KLOE has performed a fit to the proper-time distribution for 
$\kl\to3\po$ decays, which can be isolated with high purity and high and 
uniform efficiency \cite{KLOE+05:KLlife}.
The fit to the distribution for \ab10 million $\kl\to3\po$ decays
(400~\Lpb) in the proper-time interval from 6 to 25~ns gives 
$\tau_{\kl}=50.92\pm0.30$~ns.
The fractional statistical error is 
\ab$10.5/\sqrt N \approx 0.33$\%. The stated error on $\tau_{\kl}$ includes
a systematic uncertainty of 0.5\%.

If, as at a \f\ factory, the total number of \kl s created is known, 
the lifetime can also be obtained from the number of decays $N_{\rm D}$ 
in a time interval $\Delta$ beginning at time $t$.
The fractional error is given by
\begin{displaymath}
\frac{\delta\tau}{\tau} =
\frac{\delta\Gamma}{\Gamma} \approx 
\frac{1}{\sqrt{N_{\rm D}}}\:\left|\frac{1 - e^{-\Gamma\Delta}}
{(\Gamma t + \Gamma\Delta)e^{-\Gamma\Delta} - \Gamma t}
\right|.
\end{displaymath}
For KLOE, $\Gamma t \approx 0.12$ and $\Gamma\Delta \approx 0.37$, so
$\delta\tau/\tau \approx 1.4/\sqrt{N_{\rm D}}$.
Using this method, KLOE finds $\tau_{\kl}=50.72\plm0.36$~ns 
(see \Sec{sec:KLBR}).
Because the two determinations are almost entirely uncorrelated and the errors
are dominated by different systematic uncertainties, the two values can be 
averaged. KLOE obtains $\tau_{\kl}=50.84\pm0.23$, which is a factor 
of \ab2 improvement over the 1972 measurement of Vosburgh et al.\ \cite{vos}.

\FIG\depmiss   \TAB\klbr
\subsubsection{$K_L$ Branching Ratios}
\label{sec:KLBR}
The KLOE measurements of the semileptonic \kl\ BRs 
\cite{KLOE+06:KLBR} are based 
on two points. The first, unique to \DAF\ and KLOE, is the use of tagging
to obtain absolute BRs; the presence
of a \kl\ is tagged by observation of a $\ks\to\pic$ decay. 
The second point is that just four modes, $\pi^\pm e^\mp\nu$ ($K_{e3}$), 
$\pi^\pm\mu^\mp\nu$ ($K_{\mu3}$), \pic\po, and 3\po, account for more 
than 99.5\% of all decays. For the first three modes, two
tracks are observed in the DC, whereas for the $3\po$ mode, only photons 
appear in the final state. The analysis of two-track and all-neutral-particle
events is quite different. The main experimental problem is ensuring that
the tagging is bias free. Because both the trigger and 
reconstruction efficiencies for $\ks\to\pic$ decays exhibit some dependence
on the \kl\ decay mode to be identified, most of the analysis revolves
around minimizing and correcting the tag bias. 

To deal properly with this problem, two more possibilities for the behavior 
of the \kl\ must be included: The \kl\ may reach the calorimeter and
interact, or it may escape the detector altogether. We define
the tag bias $B_i$ for \kl\ channel $i$ as one minus the ratio of the \ks\
detection efficiency when $\kl\to i$ to the \ks\ detection efficiency 
averaged over all possible \kl\ ``channels,'' including the last two
listed above. The tag bias is minimized by retaining only events in which 
it is possible to verify that the $\ks\to\pic$ decay satisfied the
calorimeter trigger by itself. MC simulations
give $B_i\approx0.03$ for the two-track \kl\ decays, and $B_i\approx 0.00$
for $3\po$. A correction of $+0.2$\% to the average tagging efficiency 
is obtained from data. An additional correction of approximately $-0.5$\%, 
also obtained from data, accounts for decreased reconstruction efficiency
for the tracks from $\ks\to\pic$, owing to the presence of the \kl\ decay 
products. This last correction depends on the \DAF\ operating conditions
and is obtained separately for different sets of runs.

Two-track events
are assigned to the three channels of interest by use of a single variable:
the smaller absolute value of the two possible values of 
$\Delta_{\mu\pi}=|{\bf p}_{\,\rm
miss}|-E_{\,\rm miss}$, where
${\bf p}_{\,\rm miss}$ and $E_{\,\rm miss}$ are the missing momentum and
energy in the \kl\ decay, respectively, and are evaluated 
assuming the decay particles are a
pion and a muon. 
Figure \hyref{\depmiss} (left panel) shows an example of a $\Delta_{\mu\pi}$
distribution. A total of approximately 13 million tagged \kl\ 
decays (328~\Lpb) are used for the measurement of the BRs. The 
numbers of $K_{e3}$, $K_{\mu3}$, and $\pic\po$ decays are obtained 
separately for each of 14 run periods by fitting
the $\Delta_{\mu\pi}$ distribution with the corresponding MC-predicted
shapes.
The signal extraction procedure is tested using
particle-identification variables from the calorimeter. Figure \hyref{\depmiss} 
(right panel) shows the $\Delta_{e\pi}$ spectrum for events with 
identified electrons, together with the results of a fit using MC shapes.
The $K_{e3(\gamma)}$ radiative tail is clearly evident.
The inclusion of radiative processes in the simulation
is necessary to obtain an acceptable fit, as well
as to properly estimate the fully inclusive radiative rates.
\\\figboxc NS56pre_fig05;14;
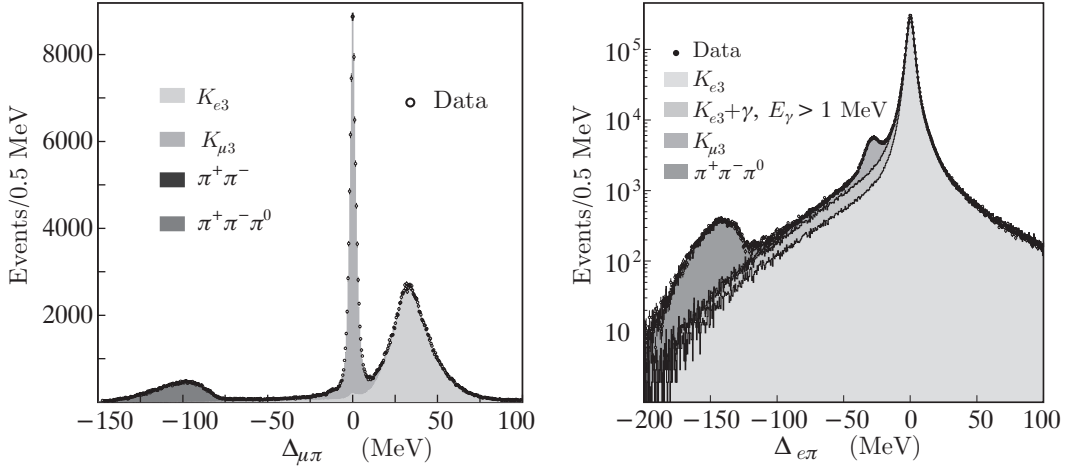
\allcap\depmiss;({\it Left panel}\/) Distribution of $\Delta_{\mu\pi}$ for a run set. ({\it Right panel}\/) Distribution of $\Delta_{e\pi}$ for events with an 
identified electron, for the entire data set.;

The decay $\kl\to3\po$ is easier to identify. Detection
of $\ge3$ photons originating at the same point is accomplished with
very high efficiency (99\%) and very little background (1.1\%).
Changing the required number of photons from
three to five reduces the background to 0.1\%, with a detection
efficiency of 88\%.
This variation in the selection criteria changes the result by 1\%.
This entire change is taken as the magnitude of the systematic 
uncertainty on the corresponding BR.

To conveniently estimate the individual efficiencies, the
BRs are obtained using the \kl\ lifetime measurement of Vosburgh
et al.\ \cite{vos}: $\tau_{\kl}=51.54\pm0.44$~ns.
The errors on the absolute BRs are dominated by the 
uncertainty on the value of $\tau_{\kl}$, which enters into the 
calculation of the geometrical efficiency. This source of uncertainty
can be all but removed (at the cost of correlating the errors among
the BR measurements) by applying the 
constraint that the \kl\ BRs must sum to unity.
The sum of the four BRs, plus the sum of the Particle Data Group
(PDG) values \cite{PDG04} 
for \kl\ decays to \pic, \pio, and $\gam\gam$ ($\sum$=0.0036), 
is $1.0104\pm0.0018\pm0.0074$.\footnote
{Throughout this article, whenever two errors are quoted on a measured
quantity, the first error is statistical and the second is systematic,
unless otherwise indicated.}
Applying the constraint gives 
the results in Table \hyref{\klbr}.\vglue2mm
\begin{center}
\cl{Table \klbr. KLOE measurements of \kl\ branching ratios}\vglue2mm
\begin{tabular}{@{}lcccc@{}}
\hline\hline
\vst Mode & BR & $\delta$ stat & $\delta$ syst-stat & $\delta$ syst \\ 
\hline
\vst\eiii     &  0.4007  &  0.0005  &  0.0004  & 0.0014 \\
\vst\muiii    &  0.2698  &  0.0005  &  0.0004  & 0.0014 \\
\vst\pio\po   &  0.1997  &  0.0003  &  0.0004  & 0.0019 \\
\vst\pic\po   &  0.1263  &  0.0004  &  0.0003  & 0.0011 \\
\hline
\label{tab:finalbr}
\end{tabular}
\end{center}\hypertarget{\klbr}\kern0pt\vglue2mm\noindent
Constraining the BRs to sum to unity and solving for the
geometrical efficiency is equivalent to determining $\tau_{\kl}$ by
the second method described in \Sec{sec:KLlife}. KLOE finds
$\tau_{\kl} = 50.72\pm0.17\pm0.33$~ns.

For the ratio $\Gamma(K_{\mu3})/\Gamma(K_{e3})$,
KLOE obtains $R_{\mu e}=0.6734\pm0.0059.$
A value for comparison can be computed from the
slope of the $f_0$ form factor.
The average of the measurements of $\lambda_0$ 
from the KTeV experiment at Fermilab \cite{KTeV+04:FF} 
and the ISTRA+ experiment at Protvino \cite{ISTRA+04:Kmu3} gives 
$R_{\mu e}=0.6640\pm0.0040$.
For the ratio $\Gamma(\kl\to\pio\po)/\Gamma(\kl\to\pic\po)$, KLOE obtains
$R_{3\pi}=1.582\pm 0.027$.
A value for comparison, $R_{3\pi}=1.579$, has been obtained from the isospin 
amplitudes derived from BRs
and Dalitz-plot slopes for $K\to3\pi$ decays \cite{binenz}.

\FIG\lalap      \def\fig#1;{Figure \hyref{#1}}
\subsubsection{The Vector Form Factor $\tilde f_+(t)$ for $K_{e3}$ Decays}
As discussed in \Sec{sec:vusl3}, knowledge of the form factor is necessary
to compute the phase-space integral that appears in \Eq{eq:Gamsl}.
This measurement is particularly delicate; it took more
than 50 years to obtain reliable results. One reason is that in the
expression for the pion energy spectrum, 
the form factor is multiplied by the kinematic density of the 
modulus-squared matrix element
$|\Ma|^2$. It so happens that the kinematic density vanishes for 
$t=t_{\rm max}$, precisely where the form factor itself is maximal. 
Therefore, the effect of the form factor on the shape of the 
pion spectrum is very small, as seen from \fig\lalap; (left panel).
The right panel of \fig\lalap; compares KLOE results for 
$\lambda_+'$ and $\lambda_+''$ \cite{KLOE+06:KLe3FF}
with those from other experiments 
\cite{KTeV+04:FF,ISTRA+04:Ke3,ISTRA+04:Kmu3,NA48+04:Ke3FF}.
\\\figboxc NS56pre_fig06;10;
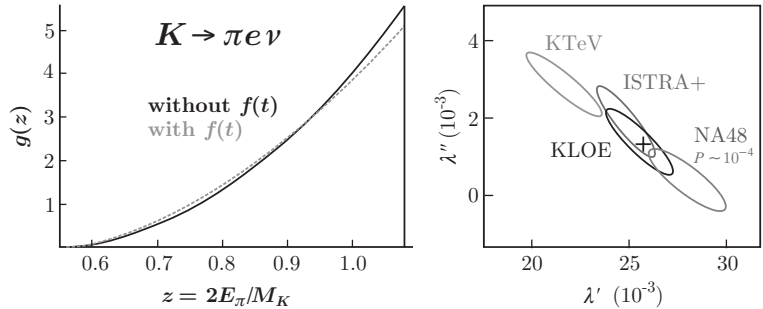
\allcap\lalap;({\it Left panel}\/) Normalized pion spectra for $K\to\pi e\nu$ decays, with $\lambda_+'=\lambda_+''=0$ and $\lambda_+'=0.0221,\ \lambda_+''=0.0023$.  ({\it Right panel}\/) KLOE $1\sig$ contours for $\lambda_+'$ and $\lambda_+''$, compared with other results. The cross gives the KLOE values obtained for a pole fit.;\noindent
The KLOE results for the slope and curvature parameters are
$\lambda_+'=\SN{(25.5\pm 1.5\pm 1.0)}{-3}$ and 
$\lambda_+''=\SN{(1.4\pm0.7\pm0.4)}{-3}$,
respectively,
with $\chi^2/{\rm ndf}=325/362$ ($P(\chi^2)=91.9$\%). 
The data have also been fit using the one-pole parameterization
(\Eq{eq:pole}). The result is
$M_V = 870 \pm 6 \pm 7$~\MeVcc; the fit gives $\chi^2/{\rm ndf}=326/363$
($P(\chi^2)=92.4$\%).

\subsection{$K_S$ Decays}
\label{sec:ksdec}
The possibility of tagging a pure \ks\ beam is unique to a \f\ factory.
At KLOE, a particularly clean tag for \ks\ decays is obtained by observing
the interaction of a \kl\ in the EMC, referred to as a ``\kl\ crash.''
The \kl\ crash is recognized as an isolated, high-energy
(typically, $E>100$~MeV) cluster that arrives
roughly 24~ns after the clusters from the \ks\ decay, as expected
for a neutral particle with $\beta \approx 0.2$. The tagging 
efficiency is approximately 30\% and is dominated by the probability for the 
\kl\ to reach the calorimeter. The position of the \kl\ crash,
together with the kinematics of the $\f\to\ks\kl$ decay, 
determines the trajectory of the \ks\
with a momentum resolution of approximately 1~\MeVc\ and an angular resolution
of better than 1\deg. The simulation of the EMC response
to the \kl\ crash in the KLOE MC has been adjusted carefully with 
reference to data \cite{Spa04}.
KLOE has used the \kl-crash tag to obtain \ks\ BR
measurements spanning six orders of magnitude. 

\subsubsection{$K_S\rightarrow \pi^+\pi^-(\gamma),\ \pi^0\pi^0$}
\label{sec:kspipi}
The two channels \pic\ and \pio\ comprise \ab99.9\% of all \ks\ decays.
The rates for the $\ks\to\pi\pi$ decays,
together with the much slower rate for $\kpm\to\pi^\pm\pi^0$,
first suggested the empirical rule that $\Delta I =1/2$ transitions are 
much favored over $\Delta I = 3/2$ transitions.
This is true in all $|\Delta S|=1$ transitions; the phenomenon remains 
poorly understood.

The ratio $\Rp\equiv\Gamma(\ks\to\pic(\gam))/\Gamma(\ks\to\pio)$ 
is a fundamental parameter of the \ks\ meson.
With almost no corrections, its measurement provides the BRs
for the dominant \ks\ decays to \pio\ and $\pic(\gam)$.
The latter decay is a convenient normalization reference for all other 
\ks\ decays to charged particles (see \Sec{sec:kse3}). 
\Rp\ is also used in the extraction of values for phenomenological parameters
of the kaon system, such as the differences in magnitude and phase of the 
$I=0,\,2$ $\pi\pi$ scattering amplitudes (see, e.g., \Ref{C+04:Kppiso}).
Finally, \Rp\ enters in the double ratio of \Eq{eq:rr}, 
which measures direct \CP\ violation in $K\to\pi\pi$ transitions.

A first KLOE measurement of \Rp\ based on 17~pb$^{-1}$ of data from 2000,
gave the result $\Rp = 2.236\pm0.003\pm0.015$ \cite{KLOE+02:KS2pi}.
This result increased the PDG average for \Rp\ by 0.028,
and correspondingly changed the PDG fit values for 
${\rm BR}(\ks\to\pic)$ and ${\rm BR}(\ks\to\pio)$ by $+0.5\%$ and 
$-1.1\%$, respectively \cite{PDG04}.
The $1.3\sigma$ difference between the KLOE value and older values 
is believed to arise from the imprecise treatment of the contribution from
$\pic\gam$ in the older measurements.
KLOE has nonzero acceptance over the entire range of photon energies
for $\ks\to\pic\gam$ decays, and the KLOE measurement of $\Rp$
is fully inclusive.
The analysis has recently been repeated using 410~\Lpb\ of 
2001--2002 data \cite{KLOE+06:KS2pi}. 
The value obtained for \Rp\ is $2.2555\pm0.0024\pm0.0050$.
Because the uncertainty on this result is dominated entirely by 
experimental systematics, the increase in statistics provided by
the 2001--2002 data is used to obtain highly accurate corrections
directly from the data itself. This, together with various 
improvements to the analysis, leads to the significant reduction
in the systematic error. The new result is consistent with the 
previous KLOE value; the overall error has been reduced to 0.25\%.
Combining the two results with attention to the common systematic errors
gives $\Rp=2.2549 \pm 0.0054$.

\FIG\kslep     \TAB\ksdata
\subsubsection{$K_S\rightarrow \pi e\nu(\gamma)$}
\label{sec:kse3}
Using the \kl-crash tag, KLOE has isolated a very pure sample of \ab13,000
semileptonic \ks\ decays and accurately measured the BRs
for $\ks\to\pi^+e^-\bar\nu(\gam)$ and $\ks\to\pi^-e^+\nu(\gam)$.

Discrete symmetries can be tested by comparing the charge asymmetries 
$A_S$ and $A_L$, defined as the difference between the widths for 
\ks\ or \kl\ decays to final states of each lepton charge, divided 
by their sum.
$A_L$ is known to an absolute accuracy of 
$\mathcal{O}(10^{-4})$ \cite{KLasym}, 
whereas $A_S$ has never previously been measured.
If \CPT\ invariance holds, the two charge asymmetries are equal
to $2\,\Re{\eps}\approx \SN{3}{-3}$.
A difference between $A_S$ and $A_L$ signals \CPT\ violation either
in the mass matrix or in the $\Delta S\neq\Delta Q$ decay amplitudes:
\begin{equation}
A_S - A_L = 4\,(\Re{\delta} + \Re{x_-}).
\label{eq:rexm}
\end{equation}
The parameter $\delta$ is defined in \Eqs{eq:kslcpt}, \ref{eq:kslncpt}, and 
\ref{eq:epsdel}.
$x_-$ is the ratio of the \CPT-odd, $\Delta S \neq \Delta Q$ decay amplitude
to the \CPT-even, $\Delta S = \Delta Q$ decay amplitude.
The most precise test of \CPT\ invariance in \ko\kob\ mixing 
comes from the CPLEAR experiment at CERN \cite{CPLEAR_redelta:98}. 
They find $\Re{\delta}$ and $\Re{x_-}$ consistent with zero, 
with sensitivities of \SN{3}{-4} and 10\up{-2}. 
The sum of the asymmetries gives
\begin{equation}
A_S + A_L = 4\,(\Re{\eps} - \Re{y}),
\label{eq:rey}
\end{equation}
where $y$ is the ratio of \CPT-odd to \CPT-even $\Delta S = \Delta Q$
decay amplitudes.
The value of $\Re{y}$ from unitarity \cite{CPLEAR:bell} is
zero to within $\SN{3}{-3}$.

From the knowledge of the semileptonic partial widths for both the 
\ks\ and \kl, it is possible to test the validity of the
$\Delta S = \Delta Q$ rule under \CPT\ invariance. With $x_+$ as 
the ratio of the \CPT-conserving amplitudes for $\Delta S \neq \Delta Q$ and 
$\Delta S = \Delta Q$ transitions,
\begin{equation}
\Re{x_+} = {1\over2}\:\frac{\Gamma(\ks\to\pi e\nu)-\Gamma(\kl\to\pi e\nu)}
{\Gamma(\ks\to\pi e\nu)+\Gamma(\kl\to\pi e\nu)}.
\label{eq:rex}
\end{equation}
In the Standard Model, a finite value of $\Re{x_+}$ on the order of 
$G_{\rm F} m_{\pi}^2 \sim 10^{-7}$ arises from second-order weak processes.
Previously, the most precise test of the $\Delta S=\Delta Q$ rule was
obtained from a study of strangeness-tagged semileptonic kaon decays by 
CPLEAR \cite{CPLEAR_rex:98}.
CPLEAR finds $\Re{x_+}$ compatible with zero to within \SN{6}{-3}.
The most precise previous measurement of ${\rm BR}(\ks\to\pi e\nu)$
was obtained by KLOE using the data from 2000 (17~pb$^{-1}$) and had a
fractional uncertainty of 5.4\% \cite{KLOE+02:KSe3}. 
The 2001--2002 KLOE data set is \ab20 times larger, and the
purity of the semileptonic decay sample has been vastly improved 
\cite{KLOE+06:KSe3}.

We outline briefly the basic steps in the analysis.
\ks\ decays are tagged by the \kl\ crash.
A cut on the $\pi\pi$ invariant
mass removes 95\% of the $\ks\to\pic$ decays and reduces the 
background-to-signal ratio to \ab80:1. Several geometrical cuts 
further improve the purity of the sample and, in particular, remove
contamination by events with early $\pi\to\mu\nu$ decays.
Finally, stringent requirements are imposed on the particle time
of flight (TOF), which very effectively separates electrons from pions and
muons and allows charge assignment of the final state.
\fig\kslep; shows the signal peak and the residual background in the
distribution of $\Delta_{Ep} = E_{\rm miss}-|{\bf p}_{\rm miss}|$
for the $\pi^-e^+\nu$ channel, where $E_{\rm miss}$ and ${\bf p}_{\rm miss}$
are respectively the missing energy and momentum at the vertex, evaluated 
in the signal hypothesis. For signal events, the missing particle is a
neutrino and $\Delta_{Ep}=0$.
\\\figboxc  NS56pre_fig07;9;
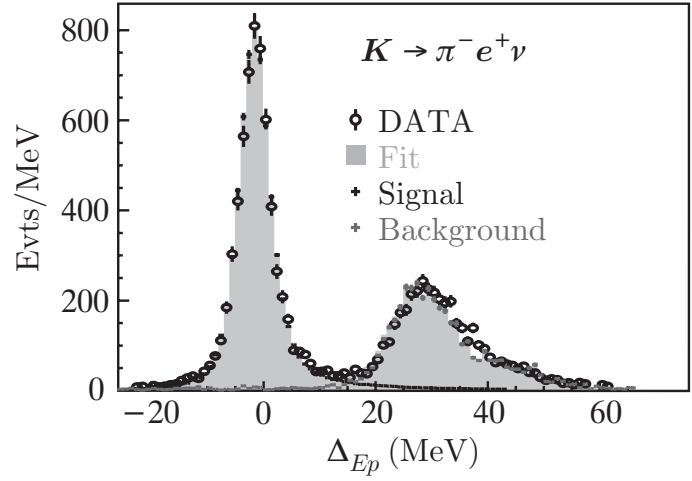
\allcap\kslep;{$\Delta_{Ep}$ distribution for $\ks\to\pi^-e^+\nu$ candidates, showing signal and background.};

The numbers of $\pi e\nu$ decays for each charge state are normalized to 
the number of \pic\ events observed, resulting in the ratios in the first 
column of Table \hyref{\ksdata}.
These ratios give the first measurement of the 
semileptonic charge asymmetry for the \ks:
\begin{displaymath}
A_S = \SN{(1.5\pm9.6\pm2.9)}{-3}.
\end{displaymath}
Using the result for \Rp\ of \Sec{sec:kspipi} (also in Table \hyref{\ksdata}), 
the absolute BRs for $\ks\to \pi\pi$ and $\ks\to\pi e\nu$
in the second column of the table are obtained.\vglue2mm
\begin{center}
\cl{Table \ksdata. KLOE measurements of $K_S$ branching ratios}\vglue2mm
\begin{tabular}{@{}lcc@{}}
\hline\hline
\vst Decay mode & BR(mode)/BR(\pic) & BR(mode) \\
\hline
\vsta $\pic$ & --- & ($69.196\pm0.024\pm0.045$)\% \\
\vst $\pio$ & 1/($2.2549\pm0.0054$) & ($30.687\pm0.024\pm0.045$)\% \\
\vst $\pi^-e^+\nu$ & \SN{5.099\pm0.082\pm0.039}{-4} & \SN{3.528\pm0.057\pm0.027}{-4} \\
\vst $\pi^+e^-\bar{\nu}$ & \SN{5.083\pm0.073\pm0.042}{-4} & \SN{3.517\pm0.050\pm0.029}{-4} \\
\vst $\pi e \nu$ & \SN{10.19\pm0.11\pm0.07}{-4} & \SN{7.046\pm0.076\pm0.051}{-4}\\
\hline
\end{tabular}
\end{center}\hypertarget{\ksdata}\kern0pt\vglue2mm\noindent
With $\tau_{\ks}$ from the PDG \cite{PDG04} and $\tau_{\kl}$ and 
BR($\kl\to\pi e\nu$) from KLOE,
\Eq{eq:rex} gives
$\Re{x_+} = \SN{(-0.5\pm3.1\pm1.8)}{-3}$,
which is more precise than the CPLEAR value \cite{CPLEAR_rex:98}
by nearly a factor of two.
Using \Eq{eq:rexm} to combine the KLOE result for $A_S$ with the 
PDG value for $A_L$
and with the CPLEAR value for $\Re{\delta}$ \cite{CPLEAR_redelta:98} 
gives
$\Re{x_-} = \SN{(-0.8\pm2.4\pm0.7)}{-3}$;
the uncertainty on this quantity is reduced by a factor of 10.
$A_S$ and $A_L$, together with the PDG value for $\Re{\eps}$
obtained without assuming \CPT\ invariance
$(|\eps|\cos{\phi_{\eps}}=(1.62\pm0.04)\times10^{-3})$,
give $\Re{y} = \SN{(0.4\pm2.4\pm0.7)}{-3}$ (\Eq{eq:rey}),
which is comparable in precision to the value obtained by 
CPLEAR using the unitarity relation \cite{CPLEAR:bell}.

\FIG\sigbox
\subsubsection{$\ks\to3\po$}
The decay $\ks\to3\po$ is purely \CP\ violating.
If \CPT\ is conserved, the BR for this decay can be 
predicted from 
$\Gamma_S = \Gamma_L|\eps + \eps'_{000}|^2$, giving 
${\rm BR} \approx \SN{1.9}{-9}$.
In KLOE, the signature is an event with 
a \kl\ crash, six photon clusters, and no tracks from the interaction
point. Background is mainly from $\ks\to\pio$ events with 
two spurious clusters from splittings or accidental activity.
Signal-event candidates are counted using the distribution in the plane
of two $\chi^2$-like discriminating variables, $\zeta_3$ and $\zeta_2$.
$\zeta_3$ is the quadratic sum of the residuals between the nominal
$\pi^0$ mass, $m_{\po}$, and the
invariant masses of three photon pairs formed from the six clusters 
present.
$\zeta_2$ is based on energy and momentum conservation in the
$\phi\to\ks\kl$, $\ks\to\pio$ decay hypothesis, 
as well as on the invariant masses of two photon pairs.
$\zeta_3$ and $\zeta_2$ are evaluated with the most favorable cluster
pairing in each case.
\fig\sigbox; shows the $\zeta_2$ versus $\zeta_3$ distributions obtained
with 450~pb$^{-1}$ of 2001--2002 data, as well as the predicted distribution
from an MC sample with an effective statistics of 5.3 times that
of the data.
From the MC distribution, the number of predicted 
background counts is $3.1\pm0.8\pm0.4$; two counts in the signal box are
observed in data.
KLOE thus obtains the 90\% C.L. limit ${\rm BR} \leq \SN{1.2}{-7}$ 
\cite{KLOE+05:KS3pi0}, which is approximately a factor of six more 
stringent than the recent limit from NA48 \cite{NA48+05:KS3pi0}.
With the additional 2~fb$^{-1}$ of data from 2004--2005 and 
improvements to the analysis under development, the KLOE limit can potentially
be reduced by an additional order of magnitude.
\\\figboxc NS56pre_fig08;11;
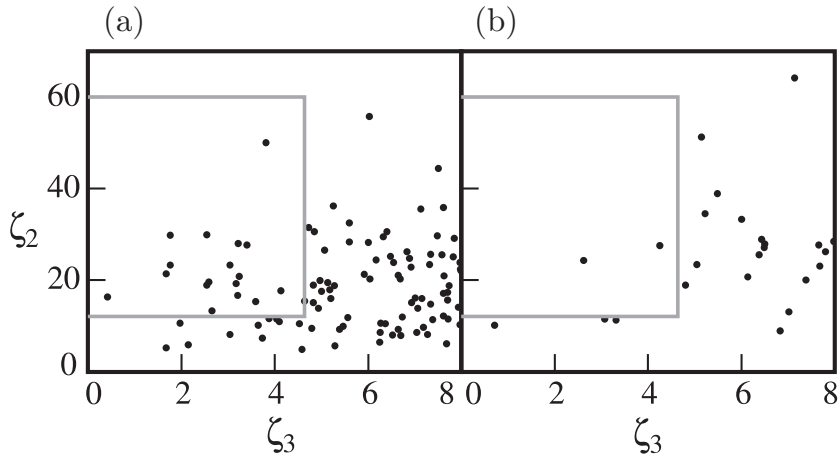
\allcap\sigbox;{Distribution of $\zeta_2$ versus $\zeta_3$ for candidate
$\ks\to3\po$ events: ({\it a}\/) from Monte Carlo simulation and
({\it b}\/) from 450~pb$^{-1}$ of KLOE data.};

\FIG\kcmu
\subsection{Charged Kaon Decays}
\label{sec:kchdec}
The $K^\pm\to\mu^\pm\nu$ ($K_{\mu2}$) decay represents approximately
two-thirds of all \kpm\ decays. It is often used as 
a reference for measuring other
BRs, and, together with the \kpm\ lifetime, provides a measurement
of the decay constant $f_K$. The most recent measurement of BR($K_{\mu2}$)
is based on 62,000 events and dates
back to 1972 \cite{ref:chiang}. The stated error is \ab0.7\%, with a
0.4\% statistical contribution. It is important to improve 
the accuracy with which this BR is known.

The measurements of the \kpm\ lifetime listed in the
PDG compilation \cite{PDG04} exhibit poor consistency.
The PDG fit has a confidence level of \pt1.5,-3,,
and the error on the recommended value is enlarged by a scale factor of 2.1.
The dominant measurement, by Ott \& Pritchard \cite{ott},
$\tau_{\kpm} = 12.380\pm0.016$~ns, was obtained by combining the results
from four runs. The systematic uncertainties for
each of the individual runs are on the order of 27~ps, yet
the overall systematic error is 14.7~ps.
This is poorly explained.
It is also curious that the stated fractional statistical error
is 0.63/$\sqrt{N}$, whereas from basic statistical considerations,
the fractional error must be at least $1/\sqrt{N}$ (and should in
fact be \ab$1.05/\sqrt{N}$ for a coverage of $6\tau_{K^\pm}$).
Thus, it is reasonable to be suspicious of the PDG value and its error.
The measurement of the absolute \kpm\ BRs and the \kpm\ lifetime
using tagged beams is an important part of the KLOE program.

\subsubsection{$K^+\rightarrow\mu^+\nu(\gamma)$}
The KLOE measurement of this BR \cite{KLOE+06:Kmu2}
is based on the use of $\km\to\mu^-\bar{\nu}$ decays
for event tagging.
Identification of a $\km\to\mu^-\bar{\nu}$ decay requires the presence of a 
two-track vertex in the DC. The nuclear interactions (NI)
of the kaons affect the BR measurement, but not the
tagging procedure.
Because the nuclear cross section
$\sig_{\rm NI}(\kp) \approx 0.01\sig_{\rm NI}(\km)$,
the use of a negative tag minimizes the corrections for NI.
(The corrections are in fact negligible.)
The large number of $K_{\mu 2}$ decays allows for a statistical 
precision of \ab0.1\%, while setting aside a generous sample for
systematic studies.

As in the analysis of the \kl\ BRs, 
to avoid any bias due to differences in the \km\ detection efficiency 
for different \kp\ decay modes, the decay products of the \km\ must
independently satisfy the trigger requirements.
The tagging efficiency exhibits a small residual dependence on the
\kp\ decay mode. This effect is corrected by MC
simulation and checked using data.

$\kp\to\mu^+\nu$ decays must have 225$\le p^*\le$400~\MeVc, 
where $p^*$ is the momentum of the charged decay particle computed
in the kaon rest frame assuming the pion mass.
The true shape of the $p^*$ distribution for signal events is obtained
from a sample of control data.
This distribution is used with the distributions for background 
sources to fit the $p^*$ spectrum (\fig\kcmu;, center panel).
\fig\kcmu; (right panel) shows the spectrum after background 
subtraction.
\\\figboxc NS56pre_fig09;14;
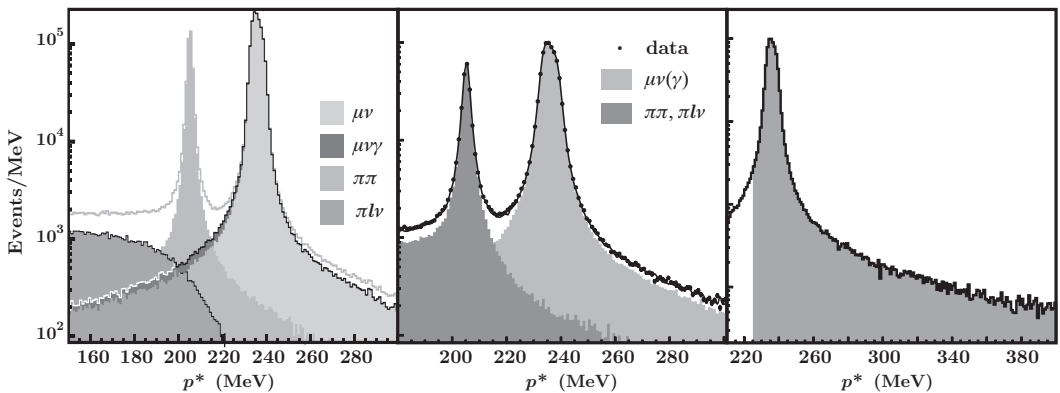
\allcap\kcmu;{({\it Left panel}\/) Monte Carlo spectra of $p^*$ for \kp\ decays, showing contributions from various channels. ({\it Center panel}\/) Distribution of $p^*$ for \ab60~\Lpb\ of 2001--2002 KLOE data. ({\it Right panel}\/) Distribution of $p^*$ after background subtraction. The shaded area is used to count $\kp\to\mu^+\nu$ events.};

In a sample of four million tagged events,
KLOE finds \ab865,000 signal events with $225 \le p^* \le 400$~\MeVc,
giving
${\rm BR}(\kp\to\mu^+\nu(\gam)) = 0.6366\pm0.0009\pm0.0015.$
This measurement is fully inclusive of 
final-state radiation (FSR) and has a 0.27\% uncertainty.

\subsubsection{Charged Kaon Lifetime}
This work is very near completion. We only mention that KLOE can measure
the decay time for individual kaons in two ways. 
The first is to obtain the proper time from the kaon path length in the DC. 
The second is based on the precise measurement of the arrival
times of the photons from $\kpm\to\pi^\pm\pi^0$ decays and the determination
of the decay point from the track vertex in the DC.
Both methods provide results accurate to the 
level of 0.1\%.

\FIG\kcsl
\subsubsection{Semileptonic Decays of Charged Kaons}
Again the measurement is based on counting decays to each channel for samples
of kaons tagged by detection of the two-body decay of the other kaon.
After transforming to the center of mass of the tagging kaon, $\mu\nu$ and 
$\pi\pi^0$ decays are distinguished because $p^*_\mu=236$~\MeVc\ and
$p^*_\pi=205$~\MeVc.
Both types of decays are used as tags.
The decay products of the tagging kaon must independently
satisfy the trigger requirements.
KLOE measures the semileptonic BRs separately for
\kp\ and \km. Therefore, 
BR($\kpm\to\po e^\pm\nu(\gam)$) and 
BR($\kpm\to\po\mu^\pm\nu(\gam)$) are each determined from four independent
measurements (\kp\ and \km\ decays; $\mu\nu$ and $\pi\po$ tags).
To identify signal events, two-body decays are first removed from the sample.
The \gam s from the \po\ are then reconstructed and provide a measurement 
of the \kpm\ decay time. Finally, the $m^2_\ell$ distribution for the 
charged secondary lepton is reconstructed by TOF.
After some further cleaning of the sample, the $m^2_\ell$ distribution
looks like that shown in \fig\kcsl;.
\\\figboxc NS56pre_fig10;11;
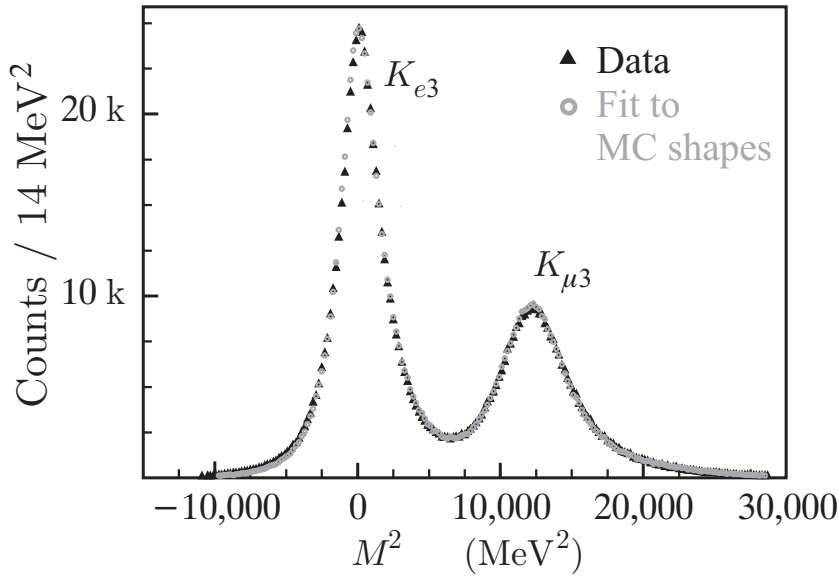
\allcap\kcsl;{Distribution of $m^2_\ell$, reconstructed using time-of-flight
information, for events selected as $K^\pm_{\ell3}$ decays. Also shown 
are the results of a fit to the Monte Carlo distributions for the two
signal channels plus residual background.};

The very good agreement of the four separate determinations of each
BR is powerful proof of the validity of the analysis and provides
an estimate of the systematic uncertainties.
The KLOE values for the BRs are
${\rm BR}(K^\pm_{e3}) = (5.047\pm0.046_{\rm stat+tag}\pm\sig_{\rm syst})\%$
and
${\rm BR}(K^\pm_{\mu3}) = (3.310\pm0.040_{\rm stat+tag}\pm\sig_{\rm syst})\%$.
These values are averages over the four different samples for each
channel and have been calculated with correlations carefully taken
into account. These results were presented at the Lisbon European
Physical Society conference
\cite{eps} and are preliminary in the sense that a careful estimate of the
systematic uncertainty on the signal selection efficiency is still
in progress.
We can however use the rms spread of the four measurements of each BR
to estimate the systematic uncertainties at
\ab0.02\% absolute for both $K_{e3}$ and $K_{\mu3}$, thus enlarging the errors
stated above to 0.05\% for both values. These results give
$R_{\mu e} \equiv \Gamma(K_{\mu3})/\Gamma(K_{e3}) = 0.656\pm0.008$.

\TAB\vusfo   \FIG\klall
\subsection{KLOE, $V_{us}$, and CKM Unitarity}
\label{sec:KLOEvus}
As noted in \Sec{sec:vus}, the KLOE data set on $K_{\ell3}$ decays
allows multiple determinations of $|V_{us}|$, with very few external 
experimental inputs.

Following the derivation in \Sec{sec:vusl3}, we compute the value
of $f^{\ko}_+(0)|V_{us}|$ from the decay rates for the five semileptonic
decay processes measured by KLOE. For the moment, we use the values
$\lambda'_+=0.0221\pm0.0011$, 
$\lambda''_+=0.0023\pm0.0004$, and
$\lambda_0=0.0154\pm0.0008$, 
obtained from a combined fit to $K_{e3}$ and $K_{\mu 3}$ results from
KTeV \cite{KTeV+04:FF} and ISTRA+ \cite{ISTRA+04:Ke3,ISTRA+04:Kmu3}.
We use the $K_S$ and $K^\pm$ lifetimes from the PDG \cite{PDG04}.
All BR measurements and the $K_L$ lifetime value are from 
KLOE, as discussed in \Secs{sec:kldec} to \ref{sec:kchdec}.
The input data and the results 
are collected in Table \hyref{\vusfo}.\vglue2mm
\begin{minipage}{\textwidth}\centering
\cl{Table \vusfo. Evaluation of $f^{\ko}_+(0)|V_{us}|$ from KLOE data}\vglue2mm
\vbox{\cl{\hbox{\renewcommand{\arraystretch}{1.1}}}}
\begin{tabular}{@{}lcccc@{}}
\hline\hline
\vsta Decay & BR\footnote{All branching ratios from KLOE.}
& $\tau$\footnote{\kl\ lifetime from KLOE; \ks\ and \kpm\ lifetimes from PDG \cite{PDG04}.}
& $\Gamma$ & $f^{\ko}_+(0)\,|V_{us}|$\footnote{Includes channel-specific $SU(2)$ 
and electromagnetic corrections.} \\
& & (ns) & ($\mu$s$^{-1}$) & \\
\hline
\vst $K_{L\,e3}$    & 0.4007(15)\footnote{Parentheses indicate errors on corresponding digits}
 & 50.84(23) & 7.88(4) & 0.2164(6) \\
\vst $K_{L\,\mu3}$  & 0.2698(15) & 50.84(23) & 5.307(32) & 0.2173(8) \\
\vst $K_{S\,e3}$ & \SN{7.05(9)}{-4} & 0.08958(6) & 7.87(10) & 0.2161(14) \\
\vst $K^\pm_{e3}$   & 0.0505(5) & 12.385(25) & 4.08(4) & 0.2178(13) \\
\vst $K^\pm_{\mu3}$ & 0.0331(5) & 12.385(25) & 2.67(4) & 0.2157(16) \\
\hline
\end{tabular}
\end{minipage}\hypertarget{\vusfo}\kern0pt\vglue2mm\noindent
Obviously, there are correlations between the different values 
for $f^{\ko}_+(0)|V_{us}|$,
particularly for the case of $K_{L\,e3}$ and $K_{L\,\mu3}$, owing to the 
non-negligible error on the \kl\ lifetime (\ab0.46\%).
With the correlations taken into account, we find
\begin{displaymath}
\langle f^{\ko}_+(0)\x|V_{us}|\rangle=0.2167\pm0.0005\kern0.75cm 
\hbox{(or \plm0.23\%),}
\end{displaymath}
with $\chi^2/{\rm ndf}=2.34/4$.
The quality of the fit is illustrated in \fig\klall;.
\\\figboxc NS56pre_fig11;9;
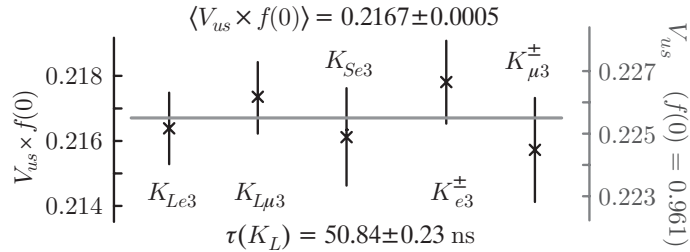
\allcap\klall;{Fit to the five KLOE values for $f^{\ko}_+(0)|V_{us}|$.};

To extract the value of $|V_{us}|$,
one needs an estimate of $f^{\ko}_+(0)$. Although the original 
calculation of Leutwyler \& Roos \cite{leutR} fell into disfavor 
a few years ago,
it appears to have been confirmed recently by vastly improved 
lattice QCD calculations \cite{lattice_ab}.
Using $f^{\ko}_+(0)=0.961\pm0.008$ \cite{leutR}, we finally obtain
\begin{displaymath}
|V_{us}|=0.2255\pm0.0019.
\end{displaymath}
To test CKM unitarity, we use 
$|V_{ud}|=0.97377\pm0.00027$ \cite{MS05}.
The first-row unitarity relation then gives
\begin{displaymath}
|V_{ud}|^2+|V_{us}|^2+|V_{ub}|^2 = 
0.9738^2+0.2255^2+{\mathcal O}(10^{-5})=0.9991\pm0.0010,
\end{displaymath}
which is in good agreement with unitarity.

Measurements of the kaon system are very promising, and we believe
that accuracies below the 0.1\% level can be achieved.
Experimental progress must be matched by better calculations,
which lattice QCD seems close to attaining.

\section{HADRONIC PHYSICS}

A \f\ factory such as \DAF\ is an ideal place at which to conduct a
number of measurements in hadronic physics. The versatility of the 
KLOE experiment is well suited for such measurements.

The leptonic widths of the \f\ meson
provide a precise test of lepton universality.
Interference between the continuum and \f-mediated amplitudes
leads to a modulation of the forward-backward asymmetry
for $e^+e^-\to e^+e^-$ events and of the cross section for 
$e^+e^-\to\mu^+\mu^-$ events,
providing sensitivity to $\Gamma_{ee}$ and
$(\Gamma_{ee}\Gamma_{\mu\mu})^{1/2}$, respectively.
From a three-point scan about the \f\ peak, conducted in 2002
(17.4~pb$^{-1}$ in total), KLOE obtains consistent values
for the two quantities, from which
$\Gamma_{\ell\ell} = 1.320\pm0.023$~keV \cite{KLOE+05:llwidth}.
 
Radiative \f\ decays ($\f\to\mbox{meson}+\gam$) are unique probes
of meson properties and structure.
The transition rates are strongly dependent on the wave function of the
final-state meson. They also depend on its flavor content, because 
the \f\ is a nearly pure $s\bar{s}$ state and because there is no 
photon-gluon coupling.

Finally, initial-state radiation (ISR) lowers the effective collision
energy from $\sqrt{s}$ to $\sqrt{s_\pi} = (s - 2E_\gamma\sqrt{s})^{1/2}$,
providing access to
hadronic states of mass $\sqrt{s_\pi}$ from threshold to $m_\f$.
KLOE has exploited ISR to measure the $\epm\to\pic$ cross section
over this entire energy range without changing the 
\DAF\ energy.

\subsection{Pseudoscalar Mesons: $\eta$ and $\eta'$}

\subsubsection{The $\eta$ Meson}
The BR for the decay $\f\to\eta\gamma$
is 1.3\%. In 2.5~fb$^{-1}$ of KLOE data, there are 100 million $\eta$ 
decays, identified clearly by their recoil against a photon of $E = 363$~MeV\@.

Current results for $m_\eta$ \cite{NA48+02:eta,GEM+05:eta}
disagree at the $8\sigma$ level.
KLOE measures $m_\eta$ using a kinematic fit to the 
$\f\to\eta\gam\to3\gam$ topology. The absolute mass scale is determined by
$m_\f$. A preliminary analysis of the 2001--2002 data
demonstrates that KLOE can measure $m_\eta$ with competitive precision
and can thereby clarify the current experimental situation.
 
Measurements of the Dalitz distributions for $\eta\to3\pi$ decays
help in constraining calculations in chiral perturbation theory.
These $\Delta I = 1$ decays proceed primarily because of strong
isospin breaking. As a result, once the phase-space integral in the
amplitude for such decays is known, the widths give access to the
ratio of quark masses $(m_s^2 - \hat{m}^2)/(m_d^2- m_u^2)$, 
with $\hat{m} = (m_u + m_d)/2$ (see \Refs{Hol02} and \citen{Bij02}).
In addition, the symmetries of the Dalitz plot for $\eta\to\pic\po$ decays
provide tests of charge-conjugation invariance. With the
2001--2002 data set, KLOE obtains a preliminary value for $\alpha$,
the Dalitz-plot slope in $\eta\to3\po$ decays, that is competitive
in precision to that from the Crystal Ball experiment \cite{CRYB+01:3pi0}, 
as well as limits on the left-right, quadrant, and sextant asymmetries 
in the Dalitz plot for $\eta\to\pic\po$ decays significantly
more stringent than the current PDG values \cite{PDG04}. 

The decay $\eta\to\po\gam\gam$ is particularly interesting in chiral
perturbation theory. As noted in \Ref{A+92:pi0gg}, there is no 
$\mathcal{O}(p^2)$ contribution, and the $\mathcal{O}(p^4)$ contribution
is small.\footnote
{We recall that in chiral perturbation theory, the effective Lagrangian
is expanded in a series of terms with increasing numbers of field 
derivatives and quark mass terms, and that the chiral orders in the 
meson sector are always even.}
The experimental value of the decay rate offers a point of 
comparison for third-order chiral perturbation theory calculations.
The history of attempts to measure the rate for this decay is confused
owing to persistent problems with background, in particular from $\eta\to3\po$.
In both existing plausible BR measurements, $\eta$ mesons
were produced via the reaction $\pi^-p\to\eta n$ \cite{GAMS+84,CRYB+05:pi0gg}.
Only the result from the Crystal Ball \cite{CRYB+05:pi0gg} is in agreement 
with chiral perturbation theory predictions.
At KLOE, the decay can be reconstructed with full kinematic closure and
without complications from certain backgrounds present in fixed-target 
experiments, such as $\pi^-p\to\pio n$. 
Using the 2001--2002 data set, KLOE has obtained the preliminary 
result ${\rm BR}(\eta\to\pi^0\gam\gam) = \SN{(8.4\pm2.7\pm1.4)}{-5}$.
This result is approximately a factor of four lower than that from the 
Crystal Ball. It is in agreement with the results of $\mathcal{O}(p^6)$
tree-level chiral perturbation theory calculations reported in
\Refs{A+92:pi0gg} and \citen{BFP96:pi0gg} and with the 
$\mathcal{O}(p^6)$ Nambu--Jona-Lasinio model calculation of \Ref{BLS96:pi0gg}.

The decay $\eta\to3\gam$ violates \C, while
the decay $\eta\to\pic$ violates both \P\ and \CP.
The Standard Model cannot accommodate either decay at any currently 
observable level.
KLOE has conducted a search for the decay $\eta\to3\gam$
using 410~pb$^{-1}$ of the 2001--2002 data \cite{eta3g}.
A clean sample of $4\gam$ events is first isolated.
Events with $\f\to\eta\gam$ and $\eta\to3\gam$ would be distinguished by
the characteristic energy of the radiated photon ($E_{\rm rad} = 363$~MeV),
which in most cases would also be the most energetic photon in the event.
No peak is observed in the distribution of the maximum photon energy for
$4\gam$ events. KLOE obtains the limit 
${\rm BR}(\eta\to3\gam)\leq\SN{1.6}{-5}$ at 90\% C.L., 
the most stringent obtained to date.
KLOE has also searched for evidence of the decay $\eta\to\pic$ 
in the tail of the $M_{\pi\pi}$ distribution for
$e^+e^-\to\pic\gam$ events in which the photon is emitted at
large polar angles ($\theta>45\deg$).
350~pb$^{-1}$ of 2001--2002 data were analyzed as described in \Sec{sec:f0}.
No peak is observed in the distribution of $M_{\pi\pi}$ in
the vicinity of $m_\eta$.
The corresponding limit is ${\rm BR}(\eta\to\pic)\leq\SN{1.3}{-5}$ at
90\% C.L. \cite{etapp}, which is more stringent than the previous limit
by a factor of 25.

\subsubsection{The $\eta'$ Meson}
The magnitude of ${\rm BR}(\f\to\eta'\gam)$
is a probe of the $s\bar{s}$ content of the $\eta'$ \cite{Ros83}. 
The ratio $\Rh \equiv {\rm BR}(\f\to\eta'\gam)/{\rm BR}(\f\to\eta\gam)$
can be related to the pseudoscalar mixing angle
$\varphi_{\rm P}$ in the basis 
$\{\ket{u\bar{u}+d\bar{d}}/\sqrt2, \ket{s\bar{s}}\}$,
offering an important point of comparison
for the description of the $\eta$-$\eta'$ mixing in extended 
chiral perturbation theory \cite{Fel00}.
At KLOE, the $\f\to\eta'\gam$ decay is identified in the 
channel in which $\eta'\to\eta\pic$ and $\eta\to\gam\gam$, whereas 
the $\f\to\eta\gam$ decay is identified in the channel in which 
$\eta\to\pic\po$.
In either case, the final state contains
a $\pi^+$, a $\pi^-$, and three photons. 
As a result, many systematics cancel 
in the measurement of \Rh.

A first KLOE measurement of this ratio was conducted using 16~pb$^{-1}$
of data from the year 2000 \cite{etapr}.
The value obtained was 
$\Rh = \SN{(4.70\pm0.47\pm0.31)}{-3}$.
This allows $\varphi_{\rm P}$ to be determined to better than 2\deg\ using
relations in \Refs{Fel00} and \citen{BES99}.
Together with additional constraints from 
the rates for $\eta'$ decays into $\rho\gam$ and $\gam\gam$, which
give information about the nonstrange quark content of the $\eta'$, 
the KLOE result for $\varphi_{\rm P}$ 
can be used to place a limit on
the gluonium content of the $\eta'$.
With $\ket{\eta'} = 
 X_{\eta'}\ket{u\bar{u}+d\bar{d}}/\sqrt2 +
 Y_{\eta'}\ket{s\bar{s}} +
 Z_{\eta'}\ket{gg}$,
where the last term represents the possible gluonium content of the 
$\eta'$ meson,
the above considerations give $Z_{\eta'}^2 = 0.06^{+0.09}_{-0.06}$,
which is compatible with zero.

A recent extension of this analysis is based on the identification 
of the $\pic7\gam$ final state (i.e., $\eta'\to\eta\pic$ with
$\eta\to 3\po$, or $\eta'\to\eta\pio$ with $\eta\to\pic\po$).
A preliminary result based on the analysis of the full
2001--2002 data set gives \Rh\
with a total error of approximately 4.5\%. Uncertainties in the intermediate
$\eta'$ BRs dominate this error. KLOE can
improve the knowledge of these BRs by using the recently
acquired data.

\FIG\kloop    \FIG\fopp
\subsection{Scalar Mesons: $f_0(980)$ and $a_0(980)$}
\label{sec:f0}
The compositions of the five scalar mesons with
masses below 1~\GeVcc\ are not well understood.
Of these, the well-established
states are the $I=0$ $f_0(980)$ and the $I=1$ $a_0(980)$.
Together with the very broad $I=0$ $f_0(\mbox{400--1200})$ ($\sigma$) 
and possible $I=1/2$ $K_0^*(800)$ ($\kappa$) 
states, these may be grouped into a nonet with an inverted mass structure.
Both the low masses of these mesons as a whole and this inverted mass 
structure are explained by the hypothesis that these
are not conventional $q\bar{q}$ mesons, but are in fact $qq\bar{q}\bar{q}$
states \cite{Jaffe}.
In this case, the mass degeneracy of the $f_0$ and $a_0$ is explained by
the fact that they contain ``hidden'' $s\bar{s}$ pairs,
whereas the $\sigma$, the state of lowest mass, is $u\bar{d}d\bar{u}$.
The quarks and antiquarks may also be paired into two pseudoscalar 
mesons \cite{WI82}.
In this case, the $f_0$ and $a_0$ are interpreted as $K\kb$ ``molecules.''
The proximity of the $f_0$ and $a_0$ to the $K\kb$
thresholds, together with the fact that these states are strongly coupled
to $K\kb$, suggests that the $f_0$ and $a_0$ must in any case be
surrounded by a cloud of virtual $K\kb$ pairs \cite{Tor95}.
Alternatively, there are unitarized quark models in which both the light
and heavy scalars are understood as dynamically generated states 
``seeded'' by bare $q\bar{q}$ states (see, e.g., \Ref{BP02}).

Decays such as $\f\to f_0\gam\to\pi\pi\gam$ 
and $\f\to a_0\gam\to\eta\po\gam$
are suppressed unless the $f_0$ and $a_0$ have significant $s\bar{s}$ content.
The BRs for these decays are estimated to be 
on the order of $10^{-4}$ if the $f_0$ and $a_0$ are $qq\bar{q}\bar{q}$
states (in which case they contain an $s\bar{s}$ pair), 
or on the order of $10^{-5}$ if the $f_0$ and $a_0$ are conventional $q\bar{q}$
states \cite{AI89,AG97}.
If the $f_0$ and $a_0$ are $K\kb$ molecules, the BR estimates
are sensitive to assumptions about the spatial extension of these 
states \cite{K+05}.
 
Apart from the BRs for these decays, the analysis of the $\pi\pi$
and $\eta\pi^0$ invariant-mass
distributions can shed light on the nature of the $f_0$ and $a_0$
because fits to the distributions provide estimates of the couplings of these
mesons to the \f\ and/or to the final-state particles.
The fit to the two-pion invariant-mass ($M_{\pi\pi}$) distribution 
for $\f\to \pi\pi\gam$ decays
may also provide evidence of a contribution from the $\sigma$.
Such fits necessarily assume a specific model for the decay mechanism.
Because of the proximity of the $f_0$ and $a_0$ masses to the $K\kb$
threshold, and because these mesons are known to couple strongly to $K\kb$,
the kaon-loop model (\fig\kloop;, left panel) is often used \cite{AI89}.
\\\figboxc NS56pre_fig12;9;
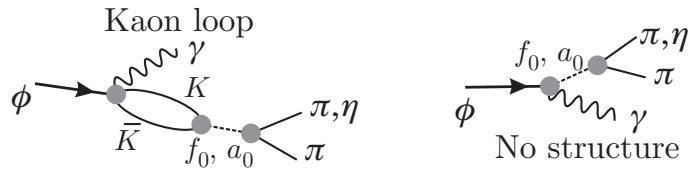
\allcap\kloop;{Amplitudes used to compute the $\pi\pi$ and $\eta\po$ mass spectra
for $\f\to S\gam$.};\noindent
The expression for the rate for E1 $\f\to S\gam$ transitions 
($S = f_0$ or $a_0$) contains a factor $E_\gam^3$ (where $E_\gam$
is the energy of the radiated photon), as required by considerations of
phase space and gauge invariance.
As a result, the invariant-mass distributions in $\f\to S\gam$ 
decays are cut off above
$m_\f$ and develop a long tail toward lower mass values.
The fit results therefore depend strongly on the scalar meson masses and 
widths.

\subsubsection{$\f\to\pi^0\pi^0\gam$ and $\f\to\eta\pi^0\gam$}
The first KLOE studies of the decays $\f\to\pio\gam$
\cite{f0} and $\f\to \eta\po\gam$ \cite{a0} were performed with the 
17~pb$^{-1}$ of data collected in 2000.

Amplitudes contributing to $\f\to\pio\gam$ include $\f\to S\gam$
(with $S = f_0$ or $\sigma$) and $\f\to\rho^0\po$ (with $\rho^0\to\po\gam$).
The $M_{\pi\pi}$ distribution was
fit with a function including terms describing the contribution from $\f\to
S\gam$, that from $\f\to\rho\pi$, and their interference. 
Two forms for the $\f\to S\gam$ term were used: one including only the
process $\f\to f_0\gam$ represented by the kaon-loop diagram of 
\fig\kloop; \cite{AI89}, and
one including an additional contribution from $\f\to\sigma\gam$, represented
by a point-like coupling.
The free parameters in the fits describing the scalars were
$m_{f_0}$,
the coupling constants $g_{f_0K^+K^-}$ and
$g_{f_0\pi^+\pi^-}$ for the kaon loop, and, 
in the fit including the $\sigma$, the coupling $g_{\f\sigma\gam}$.
The fits gave negligibly small contributions from the $\rho\pi$ 
and associated interference terms.
The fit including the $\sigma$ gave a significantly
better value of $\chi^2/{\rm ndf}$.
The value obtained for ${\rm BR}(\f\to\pio\gam)$ was 
\SN{(1.09\pm0.03\pm0.05)}{-4}, in good agreement with the results
from the CMD-2 \cite{CMD2+99:f0a0} and SND \cite{SND+00:f0} experiments
at the VEPP-2M collider in Novosibirsk, and significantly
more precise. 

The $\f\to\eta\po\gam$ channel is simpler to treat in that there is no
contribution from the $\sigma$. The amplitudes contributing are
$\f\to a_0\gam$ and $\f\to\rho^0\po$, with $\rho^0\to\eta\gam$.
In the KLOE analysis, final states corresponding to
two different $\eta$ decay channels were studied:
$\eta\to\gam\gam$ ($5\gam$ final state) and
$\eta\to\pic\po$ ($\pic5\gam$ final state).
The $M_{\eta\pi}$ distributions for each sample
were fit simultaneously with a function analogous to that discussed
above.
The fit parameters describing the $a_0$ were
the coupling constants $g_{a_0K^+K^-}$ and $g_{a_0\eta{\pi^0}}$;
the mass of the $a_0$ was fixed to the PDG value \cite{PDG04}.
The fit gave a contribution from the $\rho\pi$ term consistent with
zero.
The values obtained for ${\rm BR}(\f\to\eta\po\gam)$
using the $5\gam$ and $\pic5\gam$ samples were 
\SN{(8.51\pm0.51\pm0.57)}{-5} and
\SN{(7.96\pm0.60\pm0.40)}{-5}, respectively.
These two values are in agreement with
each other and with the results from
CMD-2 \cite{CMD2+99:f0a0} and SND \cite{SND+00:a0};
again, the KLOE data substantially improve the knowledge of these 
BRs.

In either case, the good quality of the fit establishes that the 
kaon-loop model provides a valid description of the decay process.
The BRs for $\f\to f_0\gam$ and $\f\to a_0\gam$ were 
obtained by integrating the appropriate terms of the fit functions.
Assuming dominance of the $\pi\pi$ and $\eta\po$ channels 
in $f_0$ and $a_0$ decays and that
${\rm BR}(f_0\to\pic) = 2\,{\rm BR}(f_0\to\pio)$, the values
${\rm BR}(\f\to f_0\gam)=\SN{(4.47\pm0.21)}{-4}$ and
${\rm BR}(\f\to a_0\gam)=\SN{(0.74\pm0.07)}{-4}$ are obtained.
These BRs are large,
as are the values obtained for the couplings of the scalars to $K\kb$,
especially for the $f_0$.
These results have been interpreted to support the hypothesis that
the $f_0$ and $a_0$ consist of a compact $qq\bar{q}\bar{q}$ core surrounded
by a virtual $K\kb$ cloud \cite{CT02}.
Various interpretations of the KLOE data exist, however, and 
the data have been reanalyzed by many 
authors \cite{AK03,Oll03,AK05}. Although these authors obtain different
results for 
${\rm BR}(\f \to f_0\gam)$ and the various couplings,
in general they find support 
for the $qq\bar{q}\bar{q}$ or $K\kb$ hypotheses in the KLOE data.
One reanalysis that arrives at a different conclusion 
is that of Boglione \& Pennington \cite{BP03}. They perform a $T$-matrix 
$\pi\pi$-$K\kb$ coupled-channel analysis of the data on
$\f\to\pio\gam$ from KLOE and SND, and
obtain ${\rm BR}(\f\to f_0\gam) = \SN{3.4}{-5}$, which
is smaller than the KLOE fit result by an order of magnitude.
They argue that, notwithstanding the large $\f\to K\kb$ 
coupling, the low-mass enhancement from gauge invariance
increases the importance of the $\f\to\pi\pi\gam$ coupling followed by
$\pi\pi\to\pi\pi$ rescattering. Thus, approximately 90\% of the observed
$\f\to\pio\gam$ width is from decay through the $\sigma$, rather 
than through the $f_0$. 

The analyses of the $\f\to\pio\gam$ and $\f\to\eta\po\gam$ decays
based on the 2001--2002 data set benefit from an increase in statistics
by a factor of \ab30 and are nearing completion.
Fits using two models are under development: an implementation of the 
kaon-loop model embodying some of the extensions of \Ref{AK05}, and
the no-structure model \cite{IMP06} illustrated in \fig\kloop; 
(right panel).
In this latter model, the scalar is described by a Breit-Wigner
amplitude with a mass-dependent width; the free parameters of the 
fit are $m_{f_0}$ and the couplings $g_{\f S\gam}$,
$g_{f_0\pi\pi}$ or $g_{a_0\eta\pi}$, and $g_{SK\bar{K}}$ \cite{Fla76}.
The fit includes two additional complex parameters to describe 
contributions to the amplitude from background processes.

\subsubsection{$\f\to\pi^+\pi^-\gam$}
\label{sec:f0ppg}
KLOE has recently published a study of the decay $\f\to f_0\gam\to\pic\gam$
based on the 2001--2002 data set \cite{f0ppg}.
Only a small fraction of the $e^+e^-\to\pic\gam$ events involve the 
$f_0$; the principal contributions are from events in which the photon
is from ISR or FSR.
At large values of the photon polar angle $\theta_\gam$, the ISR contribution
is strongly reduced; the analysis is performed on the $M_{\pi\pi}$ 
distribution for events with $\theta_\gam > 45\deg$.
The $M_{\pi\pi}$ distribution is fit with a function consisting of the
terms
\begin{displaymath}
\frac{d\sigma}{dM_{\pi\pi}} = 
\frac{d\sigma_{\rm ISR}}{dM_{\pi\pi}} +
\frac{d\sigma_{\rm FSR}}{dM_{\pi\pi}} +
\frac{d\sigma_{\rho\pi}}{dM_{\pi\pi}} +
\frac{d\sigma_{S\gam}}{dM_{\pi\pi}} \pm
\frac{d\sigma_{\rm int}}{dM_{\pi\pi}}.
\end{displaymath}
Analytic expressions for the ISR, FSR, and $\rho\pi$
($e^+e^-\to\rho^\pm\pi^\mp$, $\rho^\pm\to\pi^\pm\gam$) terms 
are taken from \Refs{AG97} and \citen{AG98}. 
The last two
terms describe the decay $\f\to S\gam\to\pi\pi\gam$ and its 
interference with FSR; the interference may be constructive
or destructive. The $M_{\pi\pi}$ distribution has been fit using
expressions for the scalar-mediated amplitude obtained from the kaon-loop
model, the no-structure model, and the $T$-matrix
coupled-channel framework.
\fig\fopp; (left panel) shows the $M_{\pi\pi}$ distribution, 
with the result
of the kaon-loop fit superimposed. The overall appearance of the distribution
is dominated by the radiative return to the $\rho$; the $f_0$ appears as
the peak-like structure in the region 900--1000~\MeVcc. 
\fig\fopp; (right panel) shows the data in this
mass region, with the ISR, FSR, and $\rho\pi$ terms subtracted.
\\\figboxc NS56pre_fig13;12;
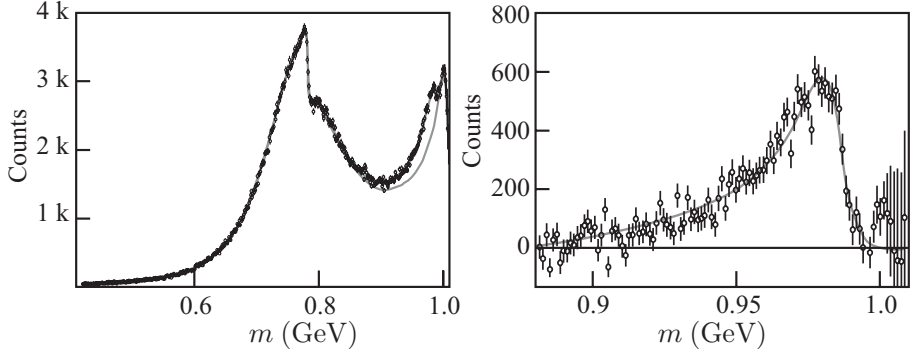
\allcap\fopp;{({\it Left panel}\/) $M_{\pi\pi}$ distribution for large-photon-angle $\pic\gam$ events from 2001--2002 KLOE data. Fits with and without the $f_0$ contribution are shown. ({\it Right panel}\/) Distribution in the region of the $f_0$, with the initial-state radiation, final-state radiation, and $\rho\pi$
contributions subtracted.};\noindent
Both the kaon-loop
and no-structure fits strongly prefer destructive interference between
the $S\gam$ and FSR amplitudes, which damps the low-mass tail of the $f_0$.
No improvement in the quality of the kaon-loop fit is observed when
the $\sigma$ is included. Both fits give values for $m_{f_0}$ 
in agreement with the PDG estimate \cite{PDG04}.
Large discrepancies between the kaon-loop and 
no-structure values are obtained for the couplings 
$g_{f_0\K^+\K^-}$ and $g_{f_0\pi^+\pi^-}$.
The kaon-loop fit gives coupling values in reasonable agreement
with those obtained from the fit to the KLOE $\f\to\pio\gam$ data
discussed above. Both fits give 
$g^2_{f_0\K^+\K^-}/g^2_{f_0\pi^+\pi^-} \approx 3$, which indicates 
that the $f_0$ couples more strongly to kaons than to pions.
Integrating the appropriate terms of the fit functions gives 
values for ${\rm BR}(\f\to f_0\gam)$ in the neighborhood of $\SN{3}{-4}$.

\subsection{The Hadronic Cross Section and $a_\mu$}

\subsubsection{The Muon Anomaly $a_\mu$}
\label{sec:amu}
The observation of the anomalous magnetic moment of the electron
helped drive the development of quantum electrodynamics (QED).
The value of the muon anomaly, $a_\mu$, 
is $(m_\mu/m_e)^2 \approx \mbox{40,000}$ times 
more sensitive than that of the electron to high-mass states
in the polarization of the vacuum. 
The Muon $g-2$ Collaboration (E821) at Brookhaven has used stored muons
to measure $a_\mu$ to 0.5~ppm \cite{E821+04}:
\begin{displaymath}
a_\mu = \frac {g_\mu - 2}{2} =  
\SN{(\mbox{116,592,080}\pm60)}{-11}.
\end{displaymath}
The muon anomaly receives contributions from QED, weak, and hadronic
loops in the photon propagator, as well as from light-by-light scattering.
The lowest-order hadronic contribution is
$a_\mu^{\rm had} =$ \ab\SN{7000}{-11}, with an uncertainty of \ab\SN{60}{-11}. 
To the extent that this uncertainty
can be reduced, the E821 measurement offers a potential probe
of new physics at TeV energy scales.

The low-energy contribution to $a_\mu^{\rm had}$ 
cannot be obtained from perturbative QCD. 
It has often been calculated from
measurements of the cross section
for $e^+e^-$ annihilation into hadrons via the dispersion integral
\begin{displaymath}
a_\mu^{\rm had} = \frac{1}{4\pi^3} \sum_f \int_{s_{\rm th}(f)}^\infty
       K(s)\, \sigma(e^+e^-\to f)\, ds,
\end{displaymath}
where $K(s)$, the QED kernel, is a monotonic function that varies
approximately as $1/s$, with $s$ the squared center-of-mass collision
energy. This amplifies the importance of the cross-section
measurements at low energy. Approximately two-thirds of the integral 
is contributed
by the process $e^+e^-\to\pic$ for $\sqrt s < 1$~GeV, i.e., in the vicinity
of $m_\rho$.
Before the arrival of the KLOE results we discuss below, calculations
of $a_\mu^{\rm had}$ from world $e^+e^-$ data
(dominated at low energies by the CMD-2 measurement of \Ref{CMD2:sighad})
led to a value of $a_\mu$ \ab$2.4\sigma$ lower \cite{DM04}
than the final value reported by E821.

Assuming the validity of the conserved-vector-current hypothesis,
the $e^+e^-$ cross section for the production of a neutral hadronic state
$f^0$ can also be obtained from the spectral function for the decay of
the $\tau$ lepton to the hadronic state $f^-$, where $f^0$ and $f^-$ are 
connected by an isospin rotation.
(The spectral function for $\tau^-\to f^-\nu_\tau$ decay
is obtained from $d\,{\rm BR}(\tau^-\to f^-\nu_\tau)/ds$, with $s$ as the 
squared mass of the state $f^-$.)
With $f^0 = \pi^+\pi^-$ and $f^-=\pi^-\pi^0$, one obtains the $I=1$ 
contribution to $\sigma_{\pic}$.
A recent evaluation \cite{D+03b} of $a_\mu^{\rm had}$ based on $\tau$ data
rather than on $e^+e^-$ data at low energies gives a value for $a_\mu$ that
is only \SN{(76\pm90)}{-11} ($0.8\sigma$)
lower than the final E821 value \cite{DM04}. Thus, even before
KLOE results became available, there was some discrepancy between 
$e^+e^-$- and $\tau$-based evaluations of $a_\mu$. We return to this 
point in \Sec{sec:taucomp}.

\FIG\sigppg
\subsubsection{Measurement of $\sigma_{\pi^+\pi^-}$ at KLOE}
\label{sec:sighad}
\DAF\ is highly optimized for running at $\sqrt s = m_\f$; $\sqrt s$
cannot be varied over a broad range.
However, ISR naturally provides access,
via the process $\epm\to\pic\gam$, to hadronic states of lower mass.
The cross section $\sigma_{\pic}$ over the entire interval 
from threshold to $m_\f$ is reflected in the
distribution of $s_\pi = M^2_{\pi\pi}$ for $\pic\gam$ events, 
\begin{equation}
\sigma_{\pic}(s_\pi) = \frac{s_\pi}{H(s_\pi|s)}\,
\frac{d\sigma_{\pic\gam}}{ds_\pi}\Big|_{\rm ISR},
\label{eq:radiator}
\end{equation}
where $H$, the ``radiator function,'' describes the ISR spectrum.
Note the subscript ISR on the differential $\pic\gam$ cross section.
This is very important because the contribution from FSR is of the same 
order as that from the ISR process. 

To correctly calculate $H$ and estimate the effects of FSR, 
an accurate simulation is critical. The KLOE analysis is based on the 
PHOKHARA event generator \cite{PHOK3}, which includes next-to-leading-order
ISR (two initial-state photons) and
has a stated accuracy of 0.5\%. Leading-order FSR and ISR-FSR terms are also
included.
In particular, ISR-FSR events are related to virtual $\pi^+\pi^-\gam$
states in the photon propagator and require special treatment
to be fully counted in the evaluation of $a_\mu^{\rm had}$.

At small photon polar angle $\theta_\gam$, ISR events are
vastly more abundant than FSR events. Requiring angular separation between
the pions and the photon further suppresses FSR \cite{BKM99}.
For these reasons, the first KLOE measurement \cite{sighad} was 
performed using events 
with $\theta_\gam < 15\deg$ and $\theta_\pi > 50\deg$ for both pion tracks.
In such events, the photon is lost down the beam pipe; $s_\pi$ 
is computed from the pion momenta.
These fiducial cuts additionally reduce background from $\f\to\pic\po$ and
$\f\to f_0\gam\to\pic\gam$ events (see \Sec{sec:f0ppg}).
They also limit the accessible range in $s_\pi$, because in $\pic\gam$ 
events with a hard photon emitted at small angle, one or both
pions from the low-mass $\pic$ pair do not enter the tracking volume.

The KLOE results obtained with 141~pb$^{-1}$ of 2001 data \cite{sighad}
are summarized in \fig\sigppg;. The left panel shows the measurement
of $d\sigma_{\pic\gam}/ds_\pi$.
\\\figboxc NS56pre_fig14;12;
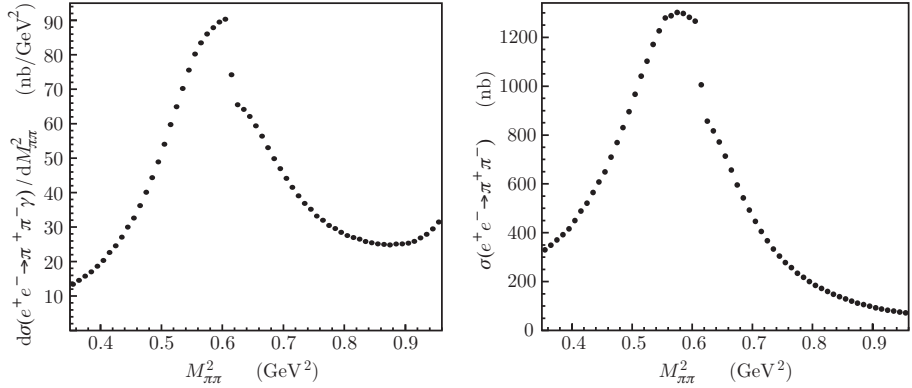
\allcap\sigppg;{({\it Left panel}\/) KLOE measurement of $d\sigma_{\pic\gam}/ds_\pi$. ({\it Right panel}\/) The cross section $\sigma_{\pic}$, including final-state radiation and without correction for vacuum polarization.};\noindent
The statistical errors range from \ab2\% at 
the lower limit in $s_\pi$ to \ab0.5\% at the $\rho$ peak.  
The experimental systematic uncertainties are mostly flat in $s_\pi$
and amount to 0.9\%.
The luminosity was estimated using large-angle Bhabha scattering and 
contributes an additional, dominantly theoretical uncertainty of 0.6\%.
The right panel of \fig\sigppg; shows the cross section 
$\sigma_{\pic}$ evaluated via \Eq{eq:radiator}. The contribution
to the systematic uncertainty on $\sigma_{\pic}$ from FSR is 
0.3\%, whereas that from the accuracy of $H$ is 0.5\%.
Finally, for the evaluation of $a_\mu^{\rm had}$, it is necessary to
remove the effects of vacuum polarization in the photon propagator
for the process $e^+e^-\to\pic$ itself by correcting for the running
of $\alpha_{\rm em}$. This contributes 0.2\% to the systematic
uncertainty.

For the contribution to $a_\mu^{\pi\pi}$ from the energy interval
$0.35 < s < 0.95$~GeV$^2$, KLOE obtains
\begin{displaymath}
a_\mu^{\pi\pi}(0.35 < s < 0.95~{\rm GeV}^2) = 
\SN{(3887 \pm 8_{\rm stat} \pm 35_{\rm syst} \pm 35_{\rm th})}{-11}.
\end{displaymath}
As a point of comparison, KLOE has evaluated the contribution to 
$a_\mu^{\pi\pi}$ for the energy interval
covered by CMD-2 ($0.37 < s < 0.93$~GeV$^2$).
The results from the two experiments differ by only \SN{30}{-11}
($0.5\sigma$).
Inclusion of the KLOE data together
with the CMD-2 data in the evaluation of $a_\mu^{\rm had}$ increases the
discrepancy between the calculated and measured values 
of $a_\mu$ to \SN{(252\pm92)}{-11} (\ab$2.7\sigma$) \cite{Hoe04}.

\FIG\pptau
\subsubsection{Comparison of $e^+e^-$ and $\tau$ Data}
\label{sec:taucomp}
As noted at the end of \Sec{sec:amu}, $e^+e^-$- and $\tau$-based evaluations
of $a_\mu^{\rm had}$ give different results. The KLOE data increase this
discrepancy, whereas new data from SND \cite{SND+05:sighad} decrease it.
\fig\pptau; compares evaluations of the pion form factor from 
$\tau$ and $e^+e^-$ data by Davier et al.\ \cite{DHZ05}.
The KLOE data clearly exhibit a trend that is systematically different
from that of the $\tau$ data.
The SND points conform
more closely to the $\tau$ data, whereas the CMD-2 points lie somewhere in
between.
As a gauge of the significance of these trends, Davier et al.\
have evaluated ${\rm BR}(\tau^-\to\pi^-\pi^0\nu_\tau)$
from the cross-section data of each of the three experiments. 
The BR values calculated from the KLOE and CMD-2 data are 
$3.8\sigma$ and $3.1\sigma$ lower than the measured value, respectively, 
and are reasonably compatible with each other. 
The BR value from the SND data is compatible with ($0.9\sigma$
lower than) the measured value.
\\\figboxc NS56pre_fig15;10;
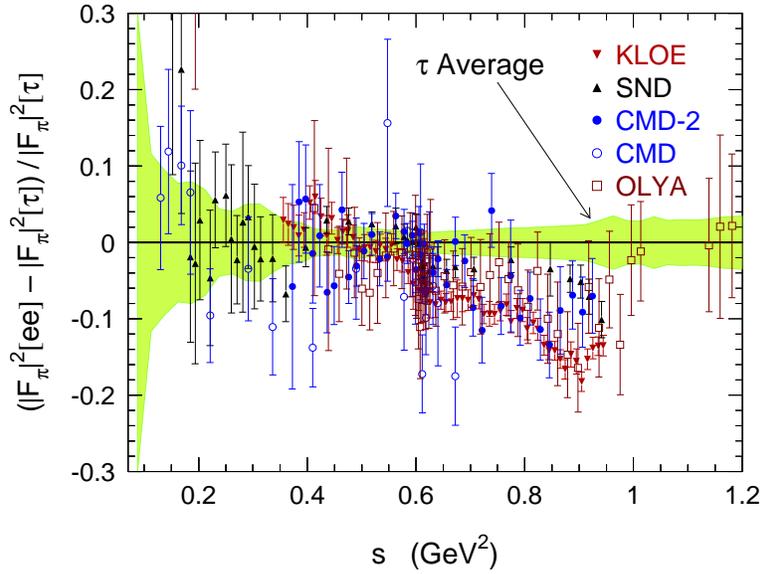
\allcap\pptau;{Relative difference between the pion form factor $|F_\pi(s)|^2$ as evaluated from $\tau$ spectral functions ({\it band shows error}\/), and $\sigma_{\pic}$ measurements from various $e^+e^-$ experiments ({\it points}\/). Reprinted figure with permission from M.~Davier, A.~H\"ocker, Z.~Zhang, {\it Rev.\ Mod.\ Phys.,}\/ accepted for publication \cite{DHZ05}. In press by the 
American Physical Society.};

Questions about possible biases in the results of any of the three
experiments may be resolved by new data and by analysis improvements.
CMD-2 has recently announced 
preliminary results from the analysis of their 1996--2000 data for 
the entire interval in $\sqrt s$ from threshold to 1.4~GeV, 
including a nearly 10-fold increase in statistics in the region covered
by their previous measurement \cite{CMD2+05:prelim}. 
KLOE prospects are discussed below.

\subsubsection{Forthcoming Improvements with KLOE}
\label{sec:sig_plans}

KLOE is actively working on repeating the small-photon-angle measurement
described above using data from 2002--2005 and a series of
improvements to the reconstruction and analysis.

The contribution to $a_\mu^{\rm had}$ from $e^+e^-\to\pic$ in the interval
in $s$ between threshold and 0.35~GeV$^2$ is approximately \SN{1000}{-11}.
As noted above, the fiducial cuts implemented in the KLOE measurement 
restrict the acceptance to higher values of $M_{\pi\pi}$. To explore the
low-mass part of the spectrum, KLOE must accept events with 
$\theta_\gam > 40\deg$.
Background from $\f\to\pic\po$ events can be rejected
by kinematic closure if explicit detection of the photon from $\pic\gam$
is required.
FSR is a more complicated problem: For $\theta_\gamma > 40\deg$, 
ISR and FSR events contribute nearly equally to the $\pic\gam$ spectrum
in the tails of the $\rho$, and the accuracy of the generator used to 
obtain FSR corrections is critical.
As discussed by Binner et al.\ \cite{BKM99}, the ISR-FSR 
interference results in a measurable charge asymmetry, which is a 
useful gauge of the accuracy of the simulation.
KLOE has performed comparisons of this type for the
analysis of $\f\to\pic\gam$ discussed in \Sec{sec:f0ppg} \cite{f0ppg}.
Moreover, with the full statistics of the complete data set,
KLOE should be able to normalize the cross section for $e^+e^-\to\pic\gam$
to that for $e^+e^-\to\mu^+\mu^-\gam$. 
This is equivalent to measuring $R$ in an energy scan and 
reduces the effects of theoretical systematics on the measurement. 
With 2~fb$^{-1}$, a $\mu^+\mu^-\gam$-normalized measurement for 
$\theta_\gam > 40\deg$ would allow the contribution to $a_\mu^{\pi\pi}$
from the interval $4m_\pi^2 < s < 0.9$~GeV$^2$ to be determined
with a statistical error of \ab\SN{10}{-11}.

In early 2006, KLOE also collected 200~\Lpb\ of data at $\sqrt s = 1000$~MeV,
at which $\sigma(\epm\to\f\to\pic\po)$ drops to 5\% of its peak value.
These data will provide a better understanding of the background 
contributions, and, by themselves, should determine the contribution 
to $a_\mu^{\pi\pi}$ as above, with a statistical error of 
\ab\SN{35}{-11} and greatly reduced systematic uncertainties.

\section{THE DEAR AND SIDDHARTA EXPERIMENTS}
\label{sec:dear}

\subsection{Kaonic Atoms and the
${\rlap{\kern.3em\raise1.9ex\hbox to.6em{\hrulefill}} K}N$ Interaction}
In the language of chiral dynamics, the meson-nucleon ``sigma terms,''
which involve matrix elements 
of the form $\bra{N}m_q\bar{q}q\ket{N}$ $(q\in{u,d,s})$,
provide a measure of the effect of the scalar quark condensate
on the nucleon, that is, the extent to which the chiral symmetry
of QCD is broken. The values of the kaon-nucleon sigma terms
are also directly related to the $\bar{s}s$ content of the nucleon.
The kaon-nucleon sigma terms are not experimentally observable,
but can be obtained indirectly from measurements of the 
$K^\pm N$ scattering amplitudes at threshold (i.e., scattering 
lengths) \cite{GS00}.
Recently, interest in the $\kb N$ interaction has been fueled
by the hypothesis that the interaction may be attractive
enough to allow the formation of deeply bound $\kb$-nuclear
states \cite{AY02}, followed by claims of evidence for the existence 
of such states (see \Sec{sec:bound}). 
Extrapolation of the existing $K^-p$ scattering data to
threshold and below is complicated by the presence of
the $\Lambda(1405)$, as well as of open channels below threshold such as
$\kb N\to Y\pi$ ($Y = \Lambda$, $\Sigma$).
Complementary information is available from the study of kaonic atoms.

A kaonic atom is formed when a $K^-$ is captured in a
high-$n$ electronic orbit of a normal atom.
The kaon undergoes a cascade of transitions down to
some final state of lower $n$ and is then absorbed
by the nucleus. The strong $K^-N$ interaction causes the energy of this
level to be shifted; the width of the spectral line is broadened by the
absorption of the $K^-$. For a light system such as hydrogen, X-ray 
transitions to the $n=1$ level can be observed.
The $1s$ level shifts ($\Delta_{1s}$) and widths ($\Gamma_{1s}$) 
of the kaonic-hydrogen atom determine
the isospin-averaged $K^-p$ scattering length $a_{K^-p}$ via the Deser
relation \cite{D+54}. 

For many years, kaonic-atom and scattering data seemed to give conflicting
results. Various analyses of $K^-p$ scattering data gave values for
$\Delta_{1s}$ and $\Gamma_{1s}$ clustered around $-360$~eV
and $510$~eV, respectively, indicating a repulsive interaction.
The few existing measurements of the X-ray spectrum of kaonic hydrogen
gave positive values for $\Delta_{1s}$, although the consistency of the
results was poor. In 1997, a kaonic-hydrogen measurement from the 
KpX experiment at KEK was reported \cite{KpX+97}.
The KpX values for $\Delta_{1s}$ and $\Gamma_{1s}$ are consistent with
those from scattering data, but significantly less precise.
The goal of the kaonic-atom program at \DAF\ 
is to measure $\Delta_{1s}$ and $\Gamma_{1s}$ at the level of a few eV.

\subsection{Measurement of the X-Ray Spectrum of Kaonic Hydrogen}

\FIG\ddra
\subsubsection{DEAR}
DEAR is literally a tabletop experiment.
The apparatus shares the
second \DAF\ interaction point with the FINUDA experiment---DEAR
can be assembled when the FINUDA experiment is retracted.

Charged kaons from \f\ decays at \DAF\ have kinetic energies of 16~MeV\@.
The DEAR target cell is mounted approximately 10~cm above the beamline at
the interaction point. Kaons traverse the thin-walled, aluminum/carbon-fiber
beam pipe and kapton entrance window, are degraded to kinetic energies
of a few MeV, and are stopped in the gaseous hydrogen (at temperature and 
pressure of 25 K and 2~bar, respectively)
of the target cell.
The target cell is 12.5~cm in diameter and 14~cm high and is constructed
with low-$Z$ materials to avoid background from X-ray fluorescence 
in the energy region of interest. The hydrogen of the target is viewed
by 16 charge-coupled device (CCD) detector chips with a total area 
of 116~cm$^2$. The pixel
size is $22.5\times22.5$ $\mu$m$^2$.
The energy resolution of the CCD chips is 
150~eV at 6~keV. In-beam energy calibration is obtained from the
fluorescence lines from high-$Z$ foils deliberately placed in the target
cell. The apparatus is further described in \Refs{DEAR+04:KN}
and \citen{DEAR+05:KH}.

After test runs and detector optimization during
2000--2001, DEAR took physics data in 2002. A first series of measurements
with the target filled with nitrogen allowed machine-background studies,
as well as an analysis of three transition lines in kaonic 
nitrogen \cite{DEAR+04:KN}. For the measurement with
hydrogen \cite{DEAR+05:KH},
58.4~pb$^{-1}$ (\ab90 million $K^-$) were collected.
In general, soft (1--10~keV) X-rays deposit energy in one or two CCD 
pixels, whereas charged particles, higher-energy photons, and neutrons 
deposit energy in several adjacent pixels. 
The X-ray spectrum was therefore formed from single- and double-pixel events.
In addition to the transition lines to be observed, the spectrum contains
a continuous background component and fluorescence lines from detector
materials. The background distribution was obtained from the analysis
of runs taken with the beams separated at the interaction point,
as well as of data from the nitrogen run. Parameters used in fits 
to the spectrum included the intensities of the 
hydrogen $K_\alpha$ ($2p\to1s$), $K_\beta$ ($3p\to1s$),
and $K_\gamma$ ($4p\to1s$) lines, the energy of
the $K_\alpha$ line, and one Lorentzian width for all $K$ lines, as the
hadronic broadening affects the $1s$ level. 
The disentanglement of the $K_\alpha$ lines from kaonic hydrogen and 
iron fluorescence, which partially overlap, required particular care.

\fig\ddra; shows a background-subtracted spectrum with the results of a fit
superimposed.
The complete analysis of the spectrum gives
$\Delta_{1s} = -193\pm37\pm6$~eV and $\Gamma_{1s} = 249\pm111\pm30$~eV.
These values are consistent with those from KpX, but with overall
uncertainties that are lower by a
factor of two. The absolute values of $\Delta_{1s}$ and $\Gamma_{1s}$
from DEAR are \ab50\% lower than those derived from
$K^-p$ scattering data. 
\\\figboxc NS56pre_fig16;7;
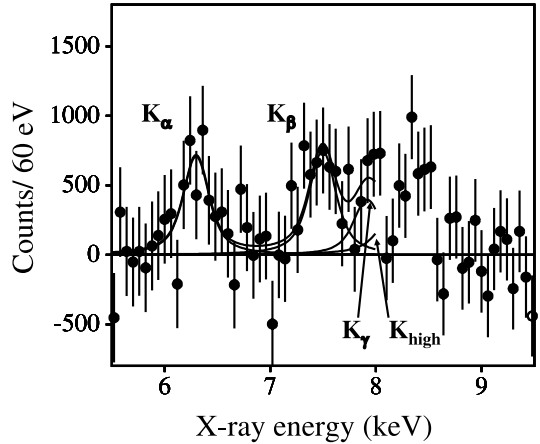
\allcap\ddra;{Background-subtracted X-ray spectrum for kaonic hydrogen
from DEAR, with fit to $K$-series spectral lines.};

\subsubsection{Interpretation}
\label{sec:KNinterp}
The $K^-p$ interaction appears to be repulsive from 
kaonic-hydrogen and scattering data. If the $\Lambda(1405)$
is interpreted as an $I=0$ $K^-p$ bound state just below threshold,
scattering through this resonance gives rise to a repulsive contribution
to the scattering length.
A systematic study of heavy-kaonic-atom data suggests that the $K^-$-nuclear
optical potential is strongly attractive in the nuclear
interior \cite{FGB94}, which may be explained by the dissolution
of the $\Lambda(1405)$ \cite{Koc94}.
Detailed calculations give variable results for the
optical potential \cite{optical}.
If the $\kb N$ attraction in medium is strong enough 
to close the decay channel to $\Sigma\pi$,
$\kb$-nuclear states with binding energies of \ab100~MeV and 
widths of a few tens of MeV may be formed \cite{AY02}, although this
scenario is not universally accepted \cite{OT05}.
Systematic analyses of the $\kb N$ interaction 
in the near-threshold region may help to establish or disprove 
this scenario; the DEAR measurement of the $K^-p$ scattering length
provides an important constraint in such analyses (see, e.g., \Ref{BNW05}).
However, the origin of the inconsistency between the 
DEAR results and the data from $K^-p$ scattering requires clarification.

\subsubsection{Future Directions: SIDDHARTA}
The program of kaonic-atom measurements at \DAF\ will be continued using
the SIDDHARTA (SIlicon Drift Detector for Hadronic Atom Research by 
Timing Applications)
apparatus \cite{Zme05}. Improvements on the precision of the
DEAR measurement require a substantial reduction in machine-background
levels.
The machine background in DEAR is primarily from beam particles
lost to intrabunch scattering or beam-gas interactions.
Effective background suppression can be obtained from timing
measurements, e.g., by requiring the coincidence between $K^+$ and $K^-$
signals in a scintillator telescope together with the signal from the
detector. The DEAR CCDs are read out once every 90~s, 
so the use of timing information is not possible.
In contrast, it is possible with an instrument based on an array of
silicon drift detectors. Silicon drift detectors are low-noise devices; the
expected energy resolution is comparable to or better than that
of the DEAR CCDs.
The SIDDHARTA detector will be ready to take data during the 2007 \DAF\
run. The planned experimental program includes 
(\textit{a}\/) the collection of 500 pb$^{-1}$ of hydrogen data 
for the measurement of $\Delta_{1s}$ and $\Gamma_{1s}$
with electron-volt precision, 
(\textit{b}\/) 1000~pb$^{-1}$ of deuterium data
to allow separate determination of the $I=0$ and $I=1$ $\kb N$
scattering lengths, and
(\textit{c}\/) a shorter run with helium that may offer
evidence in favor of deeply bound $K^-\,$$^3$He states \cite{Aka05},
if they exist.

\section{FINUDA}

\FIG\finde   \FIG\finre
\subsection{Hypernuclear Physics at a \f\ Factory}
A hypernucleus is formed when one or more nucleons in a conventional
nucleus are replaced by hyperons (most commonly $\Lambda$s).
The hyperon $(Y)$ is distinguished from the other nucleons by its strangeness
quantum number and is not subject to the Pauli exclusion principle.
It can therefore occupy any nuclear level.
The hypernuclear level structure provides valuable information about
the $YN$ interaction.
Hypernuclei with $A>5$ decay principally 
via the nonmesonic channels
$^A_\Lambda Z \to$ $^{A-2}Z\,nn$ and
$^A_\Lambda Z \to$ $^{A-2}(Z-1)\,np$,
where the underlying process is $\Lambda N \to NN$.
Thus, the decays of hypernuclei also offer information about the 
weak interaction between baryons (see \Ref{AG02} for a review).

FINUDA is a fixed-target experiment at an $e^+e^-$ collider. 
Hypernuclei are formed when $K^-$s from \f\ decays
are stopped in targets surrounding the beam pipe and interact via
$K^-_{\rm stop}\,^AZ \to$ $^A_\Lambda Z\,\pi^-$.
The spectrum of hypernuclear levels is obtained from the 
momentum distribution of the $\pi^-$ emitted when the hypernucleus is
formed at rest. In addition, hypernuclear decays can be studied 
by reconstructing the charged decay products ($\pi^-$, $p$, $d$).
The low energy of the $K^-$ ``beam'' ($E_{\rm kin} = 16$~MeV) allows
targets as thin as 0.2~g/cm$^2$ to be used, so that the intrinsic momentum
resolution of the spectrometer can be fully exploited.
FINUDA covers a solid angle
of \ab$2\pi$. Because of the large acceptance of the detector,
\ab80 hypernuclei/hour can be observed at a luminosity of $10^{32}$~\Lcms,
assuming a capture rate of $10^{-3}$.

The FINUDA detector is illustrated in \fig\finde;.
The detector is subdivided into three concentric cylindrical regions:
an interaction/target region,
an external tracking region, and an outer TOF barrel.
A superconducting solenoid 1.5~m in radius and 2.1~m in length
surrounds the entire experiment and provides a 1~T magnetic field.
\\\figboxc NS56pre_fig17;9.5;
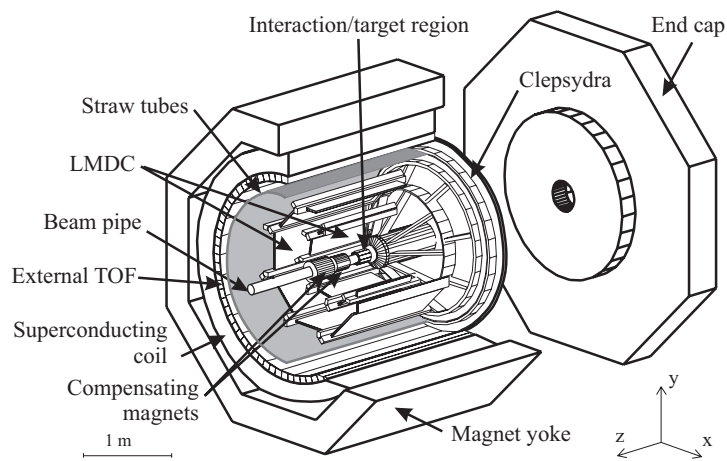
\allcap\finde;{Diagram of the FINUDA experiment.};\noindent
Kaons from \f\ decays at the interaction point traverse
the beryllium beam pipe and enter the interaction/target region.
Before encountering one of the eight solid targets arranged in an 
octagonal array about the beam pipe, they cross
an inner TOF barrel (time resolution of $\sigma_t = 250$~ps), which is 
used for trigger definition,
and a layer of double-sided silicon-microstrip detectors 
(spatial resolution of $\sigma = 30~\mu$m), which are used to reconstruct
the interaction positions in the targets.
The secondaries emerge into the external tracking region, which 
contains four concentric detector systems: a second layer of silicon 
microstrips; two cylindrical arrays of planar, low-mass DCs;
and a straw-tube detector consisting of six layers of longitudinal and stereo
tubes. The external tracking region is filled with helium to minimize 
the effects of multiple scattering.
The design momentum resolution of the tracking system for a $\pi^-$ with
$p=270$~\MeVc\ (typical of hypernuclear formation) is 0.8~\MeVc\ FWHM,
which determines the limiting energy resolution for hypernuclear levels.
The external TOF barrel consists of 72 10-cm-thick scintillator slabs.
The time resolution is 350~ps; the detection efficiency for neutrons
from hypernuclear decay is 10\%.
More information about the FINUDA detector is available in 
\Refs{FINUDA+95:tech} and \citen{FINUDA+05:LC12}.

FINUDA first took data between October 2003 and March 2004.
50 pb$^{-1}$ were collected for machine studies and detector
calibration; 200 pb$^{-1}$ were collected for analysis. 
The eight targets used were $^6$Li (isotopically enriched, two targets),
$^7$Li, $^{12}$C (three targets), $^{27}$Al, and $^{51}$V.
Much data exist on $^{12}_\Lambda$C; the three $^{12}$C targets
were included principally to provide calibration for the experiment's
maiden run. Using these targets, FINUDA has obtained a precise measurement
of the $^{12}_\Lambda$C excitation spectrum (\Sec{sec:LC12}).
The $^6$Li and $^7$Li targets allow FINUDA to observe the nonmesonic
decays of light hypernuclei.
The light targets ($^6$Li, $^7$Li, and $^{12}$C) 
can also be used to search for neutron-rich 
hypernuclei \cite{Pal04} and deeply bound 
$\kb$-nuclear states (\Sec{sec:bound}).
Using the heavier targets, FINUDA should improve 
significantly on the current precision of the excitation spectra for
$^{27}_\Lambda$Al and $^{51}_\Lambda$V, as 
well as measure the ground-state capture rates.

\subsection{First Results from the FINUDA Experiment}

\subsubsection{Reaction Spectrum of $^{12}_\Lambda$C}
\label{sec:LC12}
The FINUDA measurement of the reaction spectrum of 
$^{12}_\Lambda$C \cite{FINUDA+05:LC12} demonstrates the
potential of the experiment for hypernuclear spectroscopy.
Candidate hypernuclear events were identified by the simultaneous presence
of $dE/dx$-identified $K^+$ and $K^-$ hits in the inner detectors, together
with a high-quality track corresponding to the $\pi^-$ in the outer
detectors. 

The binding-energy spectrum of $^{12}_\Lambda$C shown in 
\fig\finre; (left panel) is obtained from the distribution of the $\pi^-$
momentum.
\\\figboxc NS56pre_fig18;11;
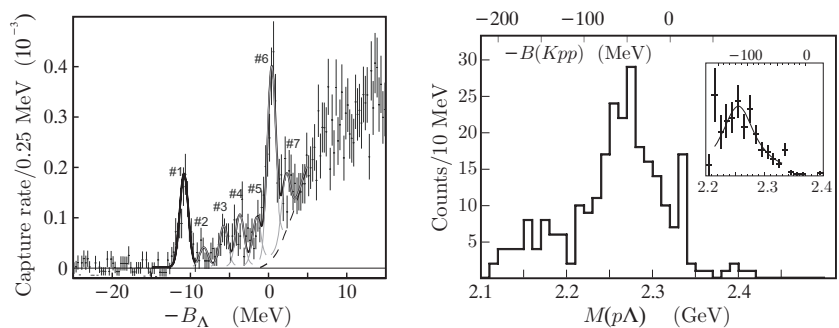
\allcap\finre;{({\it Left panel}\/) Binding-energy spectrum for $^{12}_\Lambda$C with fit to peaks plus quasi-free background ({\it dashed line}\/). ({\it Right panel}\/) $M_{\Lambda p}$ spectrum for events from light targets with $\cos\theta_{\Lambda p}<-0.8$. ({\it Right panel inset}\/) Peak after acceptance corrections. From FINUDA.};\noindent
The two prominent peaks at $B_\Lambda \approx 11$~MeV (\#1 in \fig\finre;)
and $B_\Lambda \approx 0$~MeV (\#6) correspond to the ground-state
configuration with the $\Lambda$ in the $s$ shell and to the
excited state with the $\Lambda$ in the $p$ shell, respectively.
The measured value of the capture rate for the ground state
is \SN{(1.01\pm0.11\pm0.10)}{-3} per stopped $K^-$.
By fitting the ground-state peak with a Gaussian, 
the energy resolution was determined to be 1.29~MeV FWHM\@.
These results were obtained without the detector fully
calibrated; the momentum resolution may be improved in future analyses.
FINUDA has also performed different fits to the entire spectrum. 
The fit shown in \fig\finre; was to seven Gaussian peaks with
free central values and weights; $\sigma$ for all peaks was fixed
to 0.55~MeV\@. The fit also included the quasi-free component
illustrated as the dashed line.

The $^{12}_\Lambda$C reaction spectrum was previously measured by
the E369 experiment at KEK \cite{E369+01},
in which hypernuclei were produced from a $\pi^+$ beam by
associated production ($\pi^+\,$$^AZ\to$ $^A_\Lambda Z\,K^+$). 
The energy resolution was 1.45~MeV FWHM\@.
FINUDA compares the positions and strengths of the peaks in \fig\finre;
(left panel) to those seen in the E369 data.
The momentum transfer to the $\Lambda$ in E369 is \ab350~\MeVc,
as compared with \ab250~\MeVc\ in FINUDA. As a result, the relative weights
observed for the $s_\Lambda$ and $p_\Lambda$ peaks are
expected to differ \cite{IMB90}.
Some differences are also noted in the structure around peaks 
\#3, \#4, and \#5.
Several excited states of the $^{11}$C core nucleus may contribute to the
structure in this energy region \cite{Mot98}. 

FINUDA uses the ground-state peak to normalize the capture rates for
the other peaks and compares the resulting values to 
older data on $^{12}_\Lambda$C from stopped $K^-$ experiments
and to the results of various calculations.
In general, the capture rates for the identified peaks
in the FINUDA spectrum are significantly larger than calculations predict,
whereas the FINUDA capture rates for the stronger peaks observed in older
experiments are in reasonable agreement with the older results.

\subsubsection{New States from Nuclear $K^-$ Absorption at Rest}
\label{sec:bound}
The E471 group at KEK claims to have observed two $S=-1$ tribaryon states
from $K^-$ capture at rest: $K^-_{\rm stop}\,$$^4{\rm He}\to XN_{\rm r}$,
where the recoil nucleon $N_{\rm r}$ is $p$ or $n$ \cite{K471+05:HYP03}.
The states are seen as peaks in the missing-mass distributions as
reconstructed from the TOF-measured velocity of the recoil nucleon.
A peak with a significance of $8.2\sigma$ in the distribution 
for events with a recoil proton 
is interpreted to be from the reaction $K^-\:(pnn)\to S^0\to\Sigma NN$ or
$\Sigma^-d$.
The $S^0(3115)$ is an isospin triplet state with an excitation energy 
of \ab44~MeV and a width of $<22$~MeV\@.
For events with a recoil neutron, a $3.7\sigma$ peak is observed and
attributed to $K^-\:(ppn)\to S^+\to\Sigma^+nn$ or $\Sigma^-pp$.
The $S^+(3140)$ is an isospin singlet with an excitation energy 
of \ab69~MeV and a width of $<22$~MeV\@.
If interpreted as $K^-NNN$ bound states (see \Sec{sec:KNinterp}),
the $S^0$ and $S^+$ have binding energies of 194~MeV and 169~MeV\@.
The nonrelativistic calculations of \Ref{AY02} do not predict the 
existence of the $S^0$; they find a state similar to the $S^+$, but
of higher mass. Such calculations are questionable.
An alternative explanation of the data in which the
peaks correspond to the process $K^-NN\to YN$ 
(with the remaining two nucleons spectators) 
has also been proposed \cite{OT05}.

While the E471 results are based only on TOF measurements, 
FINUDA can reconstruct the charged particles in the final state
and compute invariant masses.
FINUDA has studied the $\Lambda p$ invariant-mass distribution following 
$K^-$ capture in the light targets ($^6$Li, $^7$Li, and $^{12}$C)
\cite{FINUDA+05:Kpp}.
For a normal two-nucleon absorption process, $M_{\Lambda p}$
should peak sharply at \ab30~MeV below $m_{K^-} + 2m_p$.
Instead, the $M_{\Lambda p}$ distribution for events
with $\cos \theta_{\Lambda p} < -0.8$ is broadly peaked, as shown in 
\fig\finre; (right panel). 
A fit gives a mass of $2255^{+5}_{-6}$$^{+4}_{-3}$~MeV
and a width of $67^{+14}_{-11}$$^{+2}_{-3}$~MeV\@.
FINUDA attributes this peak to the
decay of a $K^-pp$ bound state, in which case
the binding energy is 115~MeV\@.
The $\Lambda p$ excitation energy is 201~MeV, which is large and
consistent with the large width.
Especially in light nuclei, two-nucleon absorption followed by 
rescattering effects could give rise to a peak like that
observed~\cite{M+06}.
We are skeptical of the need to invoke 
deeply bound $\kb$-nuclear or multiquark states to explain the
peaks observed by E471 and FINUDA.

FINUDA will resume data taking during summer 2006 and expects to
collect 1~fb$^{-1}$. The planned target set 
includes light targets such as $^6$Li and $^9$Be, which are nearly
$\alpha d$ and $\alpha\alpha n$ clusters, and are ideal for use
in the search for new states.

\section{THE FUTURE OF \DAF}

KLOE finished taking data at $\sqrt s = m_\f$ in December 2005 and 
collected a small amount of off-peak data
necessary to round out the data set in January--March 2006. 
During the latter half of 2006,
\DAF\ running will be dedicated to FINUDA; SIDDHARTA will run 
in 2007. Thus the original \DAF\ program will be completed, 
except for the search for direct \CP\ violation.

Discussion concerning what sort of facility may replace \DAF\
began formally at Alghero in September 2003 \cite{Alghero}.
Many options have been considered \cite{Future}.
One likely alternative is a second-generation \DAF, with a luminosity
increased by a factor of \ab5 at $\sqrt s = m_\f$, possibly capable 
of running at energies up to the $N\bN$ threshold.
However, a definite proposal will probably not emerge until late 2006.

Meanwhile, the KLOE study of leptonic and semileptonic kaon decays 
has been extremely successful. Using \ab1/5 of the available data,
KLOE has measured $f^{\ko}_+(0)|V_{us}|=0.2167\pm0.0005$, i.e., 
with an accuracy of \plm0.23\%, 
from which $|V_{us}|=0.2255\pm0.0019$ and 
$|V_{ud}|^2+|V_{us}|^2+|V_{ub}|^2=0.9991\pm0.0010$.
The error is dominated by theoretical uncertainties on $f^{\ko}_+(0)$
that will likely decrease in the near future. 
These results establish that it is possible to reach accuracies 
below 0.1\%.

The decay-rate measurements discussed provide
access to a series of effective couplings,
such as $G_e$, $G_\mu$, $G_q\,|V_{ud}|$, and $G_q\,|V_{us}|$.
If there is new physics at the TeV scale, it will be reflected in the
values of these effective couplings.
Lepton universality is the statement that $G_e = G_\mu \cdots = G_\ell$.
Unitarity in the quark sector refers to the CKM matrix;
in addition, the full Standard Model requires $G_\ell=G_q=G_{\rm F}$. 
New physics may change this simple picture at the level of 0.1\%.
In testing universality, if we limit ourselves to 
$e$, $\mu$, $u$, $d$, and $s$, we have five parameters and
need five measurements.
Kaons provide three:
$\Gamma(K_{\ell3})$, $\Gamma(K_{\mu2})$, and $\Gamma(K_{e2})$.
With the addition of the muon and pion lifetimes, we have five.
The half-lives for $0^+\to0^+$ nuclear $\beta$-transitions
provide a welcome check. We consider these measurements
central to the future of a \f\ factory. 

\vspace{3ex}
\noindent{\bf\large Acknowledgements}\\

We would like to thank our fellow KLOE collaborators
for their heroic dedication and brilliance in bringing
about {\ita ``questa bella e grande avventura.''}
We would also like to warmly congratulate
the members of the \DAF\ staff and of the DEAR and FINUDA
Collaborations on their achievements.
Finally, we thank Juliet Lee-Franzini for a critical
reading of this manuscript.

\end{document}